\documentclass[11pt]{article}
\pdfoutput=1 

\usepackage{jheppub} 

\usepackage[T1]{fontenc} 
\usepackage{blkarray}
\usepackage{relsize}

\usepackage{filecontents}

\usepackage[utf8]{inputenc}
\usepackage{amssymb, amsthm, tabto, epigraph, mathtools, dsfont, verbatim, slashed, mathrsfs}
\usepackage{graphicx, wrapfig}
\usepackage{enumitem}
\usepackage{relsize}
\usepackage{subcaption}
\usepackage{csquotes}
\usepackage{hyperref}
\usepackage{physics}
\usepackage[dvipsnames]{xcolor}
\usepackage{color}

\usepackage{amsmath}
\usepackage{empheq}
 
\setlist{nolistsep}
\hypersetup{
    colorlinks=true,
    linkcolor=Mahogany,
    filecolor=magenta,      
    urlcolor=blue,
}

\newlength\dlf  

\allowdisplaybreaks

\setlist{nolistsep}
\hypersetup{
    colorlinks=true,
    linkcolor=Maroon,
    filecolor=Maroon,      
    urlcolor=Maroon,
    citecolor=Maroon
}

\newcommand{\Mod}[1]{\ (\mathrm{mod}\ #1)}

\DeclareMathOperator{\calA}{\mathcal{A}}
\DeclareMathOperator{\calN}{\mathcal{N}}

\DeclareMathOperator{\bbR}{\mathbb{R}}

\DeclareMathOperator{\bbN}{\mathbb{N}}
\DeclareMathOperator{\calO}{\mathcal{O}}

\DeclareMathOperator{\dDisc}{\mathrm{dDisc}}
\DeclareMathOperator{\aboveContour}{\rotatebox[origin=c]{0}{$\curvearrowleft$}}
\DeclareMathOperator{\belowContour}{\rotatebox[origin=c]{180}{$\curvearrowright$}}

\usepackage{amssymb}
\usepackage{tikz}
\usetikzlibrary{shapes,arrows,cd,chains,decorations.markings,decorations.pathmorphing,calc}
\tikzset{
->-/.style args={#1rotate#2}{decoration={markings, mark=at position #1 with {\arrow[scale=1.5,rotate = #2 ]{stealth}}}, postaction={decorate}}
}
\tikzset{curve/.style={settings={#1},to path={(\tikztostart)
    .. controls ($(\tikztostart)!\pv{pos}!(\tikztotarget)!\pv{height}!270:(\tikztotarget)$)
    and ($(\tikztostart)!1-\pv{pos}!(\tikztotarget)!\pv{height}!270:(\tikztotarget)$)
    .. (\tikztotarget)\tikztonodes}},
    settings/.code={\tikzset{quiver/.cd,#1}
        \def\pv##1{\pgfkeysvalueof{/tikz/quiver/##1}}},
    quiver/.cd,pos/.initial=0.35,height/.initial=0}
\tikzset{tail reversed/.code={\pgfsetarrowsstart{tikzcd to}}}
\tikzset{2tail/.code={\pgfsetarrowsstart{Implies[reversed]}}}
\tikzset{2tail reversed/.code={\pgfsetarrowsstart{Implies}}}

\newmuskip\pFqskip
\mathchardef\pFcomma=\mathcode`, 

\newcommand*\pFq[5]{%
  \begingroup
  \begingroup\lccode`~=`,
    \lowercase{\endgroup\def~}{\pFcomma\mkern\pFqskip}%
  \mathcode`,=\string"8000
  {}_{#1}F_{#2}\biggl[\genfrac..{0pt}{}{#3}{#4};#5\biggr]%
  \endgroup
}

\newcommand*\pFqreg[5]{%
  \begingroup
  \begingroup\lccode`~=`,
    \lowercase{\endgroup\def~}{\pFcomma\mkern\pFqskip}%
  \mathcode`,=\string"8000
  {}_{#1}\tilde{F}_{#2}\biggl[\genfrac..{0pt}{}{#3}{#4};#5\biggr]%
  \endgroup
}

\graphicspath{{./images/}}

\title{Notes on Resonances and Unitarity from Celestial Amplitudes}

\abstract{We study the celestial description of the $O(N)$ sigma model in the large $N$ limit as introduced by Coleman, Jackiw and Politzer. Focusing on three-dimensions, we analyze the implications of a UV complete, all-loop order 4-point amplitude of pions in terms of correlation functions defined on the celestial circle. We find these retain many key features from the previously studied tree-level case, such as their relation to Generalized Free Field theories and crossing-symmetry, but also incorporate new properties such as IR/UV softness and S-matrix metastable states. In particular, to understand unitarity, we propose a form of the optical theorem that controls the imaginary part of the correlator based solely on the presence of these resonances. We also explicitly analyze the conformal block expansions and factorization of four-point functions into three-point functions. We find that summing over resonances is key for these factorization properties to hold. We end with some topics for future study.}

\author[1]{Diego Garc\'ia-Sep\'ulveda,}
\author[2,3,4]{Alfredo Guevara,}
\author[5,6]{Justin Kulp,}
\author[7]{and Jingxiang Wu}

\affiliation[1]{Kadanoff Center for Theoretical Physics \& Enrico Fermi Institute, University of Chicago}
\affiliation[2]{Center for the Fundamental Laws of Nature, Harvard University, Cambridge, MA 02138, USA}
\affiliation[3]{Black Hole Initiative, Harvard University, Cambridge, MA 02138, USA}
\affiliation[4]{Society of Fellows, Harvard University, Cambridge, MA 02138, USA}
\affiliation[5]{Perimeter Institute for Theoretical Physics, Waterloo, ON N2L 2Y5, Canada}
\affiliation[6]{Department of Physics \& Astronomy, University of Waterloo, Waterloo, ON N2L 3G1, Canada}
\affiliation[7]{Mathematical Institute, University of Oxford,
Andrew-Wiles Building, Woodstock Road, Oxford, OX2 6GG, UK}

\emailAdd{dgarciasepulveda@uchicago.edu}
\emailAdd{aguevaragonzalez@fas.harvard.edu}
\emailAdd{jgjkulp@gmail.com}
\emailAdd{Jingxiang.Wu@maths.ox.ac.uk}

\begin{document}

\maketitle

\section{Introduction}

The celestial holography program aims to characterize flat-space S-matrices of diverse theories in terms of boundary conformal correlators \cite{stromingerCelestial, pasterskiShao,fullVirasoro:3}, constituting a promising approach to the long-standing problem of flat-space holography. The framework has been succesful in unveiling new structures of the S-matrix, in particular regarding its infrared structure \cite{Strominger:2017zoo}, yet many questions arise regarding the nature of the so-defined celestial conformal field theories (CCFTs), see \cite{Raclariu:2021zjz, Pasterski:2021rjz,Pasterski:2021raf} for recent reviews. In particular, a precise statement of unitarity and a classification of the spectrum are missing in CCFTs, although recent progress has been made for both massive and massless S-matrices in \cite{lamShao,Nandan:2019jas,Chang:2021wvv,Melton:2021kkz,Fan:2021isc,Fan:2021pbp,Atanasov:2021cje,Guevara:2021tvr}.

Along such lines, it has been observed that well-understood analytic properties of scattering amplitudes have a novel realization in celestial correlation functions, yielding surprising features from the viewpoint of standard CFTs \cite{celestialSymmetries,celestialConstraints,Pate:2019lpp,Fotopoulos:2019tpe,Fotopoulos:2019vac,Fotopoulos:2020bqj,Fan:2019emx,Arkani-Hamed:2020gyp}. In particular, references \cite{lamShao,Nandan:2019jas,Chang:2021wvv,Atanasov:2021cje,Guevara:2021tvr,Melton:2021kkz,ShuHengTalk} have considered the celestial realization of the tree-level S-matrix with massive particle exchange at both three and four points. It was observed that the three-point correlation functions are regular and non-distributional (as opposed to the all-massless ones) and the four-point function has a simple structure closely related to a Generalized Free Field Theory (GFFT) (see also \cite{ShuHengTalk}). Indeed, using a scalar theory as an example, \cite{lamShao} pointed out that such four-point correlation functions in CCFT satisfy a novel realization of the optical theorem. More precisely, the four-point function has an imaginary piece controlled by three-point correlation functions through a conformal partial-wave expansion \cite{lamShao,Nandan:2019jas,Atanasov:2021cje,Melton:2021kkz}. This follows directly from the analytic factorization of the S-matrix and provides a first approach to the underlying unitarity of CCFTs.

A number of pressing questions arise from the previous discussion:
\begin{itemize}
    \item The optical theorem applies to the imaginary piece of the 4-point function. How does unitarity extend to the full correlation function?
    \item What is the precise relation of the partial wave expansion to the standard conformal block decomposition from the CCFT point of view? How does the CCFT spectrum emerge from the principal continuous series? 
    \item Do any of the above conclusions hold beyond tree level? Do they hold at finite coupling?
\end{itemize} 
The last point is important because, as observed in \cite{Arkani-Hamed:2020gyp}, celestial amplitudes are ``anti-Wilsonian.'' This means that unlike their bulk analogs, they are highly sensitive to UV physics, because the scattering of boost eigenstates covers the complete range of energies. 

In this work we will introduce a novel realization of unitarity that is particular to celestial CFTs as opposed to their underlying QFT. In QFT, unitarity (and generalized unitarity) requires one to sum over a stable spectrum of states \cite{VELTMAN} (see also \cite{Denner:2014zga, Donoghue:2019fcb, Menezes:2021tsj,Chang:2021wvv}. In turn, the analysis of singularities associated to bound states and metastable particles (resonances) requires a particular prescription for analytic continuation of the S-matrix, i.e \cite{Hannesdottir:2022bmo}. Here, we will show that in the CCFT context the presence of these singularities is manifest in four-point correlation functions. The choice of S-matrix prescription turns into a definition of the Mellin integral transform, which connects the momentum basis to the conformal basis. In this setup, one finds that in the conformal basis resonances completely control the imaginary part of the correlation function and resum the contribution from bound states.

The object under study is a UV/IR finite scattering amplitude valid to all loop-orders in the large $N$ limit, corresponding to the three-dimensional $O(N)$ model first studied by Coleman, Jackiw, and Politzer. It exhibits spontaneous symmetry breaking, and by introducing auxiliary fields the effective action can be summed to all orders. This allows us to explicitly derive and check the unitarity properties of the corresponding 1d CCFT non-perturbatively. This includes partial wave decomposition for the full correlator and factorization of the imaginary part. Interestingly, in this particular case the four-point amplitude retains much of the GGFT structure previously observed at tree-level by Lam and Shao \cite{lamShao}, which passes the usual unitarity constraints such as positivity, but also incorporates new relevant features of CCFT such as color structure. We will also detail the general extent to which our arguments are expected to hold, in a more diverse class of theories.






This paper is organized as follows: In Section \ref{sec:effDescON} we recount the famous solution to the $O(N)$ model described by Coleman, Jackiw, and Politzer, and use it to obtain the $\pi^a\pi^b\to\pi^c\pi^d$ and $\pi^a\pi^b\sigma$ scattering amplitudes in Section \ref{sec:threeFourAmps}, with focus on (2+1)d.

In Section \ref{sec:CelestialON} we introduce the celestial amplitude that we will study for the rest of the paper. In Section \ref{sec:celestialSetup} we recap the structure of a celestial CFT correlator and set our notation, contrasting 1d CCFTs to 2d CCFTs. We compute the actual $O(N)$ CCFT amplitude in \ref{sec:ONAmplitude} and consider the IR and UV limits in Section  \ref{sec:EFTExpansion}. In Section \ref{sec:SMatCrossing} we look at the permutation relations coming from scattering four bosons, and the implications for the (stripped) CCFT four-point function.

In Section \ref{sec:GenOpticalThm} we study how resonances manifest in celestial amplitudes via non-zero contributions to the imaginary part of the celestial amplitude, extending a result of \cite{Chang:2021wvv} to the case of resonances. We illustrate our observations in the case of the $O(N)$ model.


In Section \ref{sec:confDec} we look at various CFT data implied by the CCFT correlator. We lay out the amplitudes in Section \ref{sec:celestialAmps} then study the conformal block decompositions in Section \ref{sec:confBlock}. We verify that the same result is obtained for conformal partial waves using the Euclidean inversion formula in Section \ref{sec:CPWs}. In Section \ref{sec:factorization} we analyze the factorization of four-point function coefficients into three-point function coefficients and explain the modification to the celestial optical theorem proposed in \cite{lamShao}.

We conclude in Section \ref{sec:conclusion} and end with a broad list of open problems and conjectures for future study inspired by our work in Section \ref{sec:openProblems}.

We set our conventions for CFT in Appendix \ref{app:CFTBackgroundConventions} and list various formula for four-point functions and conformal blocks and partial waves in Appendix \ref{sec:1dCFTGeneral} and Appendix \ref{app:blocksAndWaves} respectively.

\section{Effective Description of the \texorpdfstring{$O(N)$}{O(N)} Model}\label{sec:effDescON}
In this section we review the relevant pieces in the large $N$ analysis of the $O(N)$ model as described in \cite{Coleman:1974jh} (see also the review \cite{mosheQuantumFieldTheory2003} and more recently \cite{Carmi:2018qzm}), which will also serve to establish some notation. We comment on the analytic structure of the correlation functions, and compute the three and four-point functions for use in celestial $\pi\pi\to\pi\pi$ scattering. For further details we refer the reader to the original paper \cite{Coleman:1974jh} which we follow closely. We use Euclidean signature throughout this section.

The $O(N)$ model is given by the following $O(N)$ symmetric Lagrangian of $N$ real scalar fields $\varphi^a$
\begin{equation}
	\mathcal{L}=\frac{1}{2} \partial_{\mu} \varphi^{a} \partial^{\mu} \varphi^{a}-\frac{1}{2} \mu_{0}^{2} \varphi^{a} \varphi^{a}-\frac{\lambda_{0}}{8 N}\left(\varphi^{a} \varphi^{a}\right)^{2}\,,
\end{equation}
where $\mu_0^2$ and $\lambda_{0}$ are the bare squared mass and bare coupling constant. The corresponding renormalized quantities will be denoted $\mu^2$ and $\lambda$ respectively. It is known that this model can be solved exactly in the large $N$ limit \cite{Coleman:1974jh}. We will not review that computation here, but instead we will simply list the key results needed. 

The first point to consider is that when $\lambda>0$ and $\mu^2 < 0$, $\varphi$ obtains a vacuum expectation value (vev)
\begin{equation}
	\expval{\varphi}^{2}=-2 \mu^{2} N / \lambda \label{eq:vev}\,,
\end{equation}
which spontaneously breaks the $O(N)$ symmetry to $O(N-1)$. Without loss of generality, we choose to preserve the $O(N-1)$ symmetry in the first $N-1$ components, which are our massless Goldstone bosons, and take $\varphi^N$ to have the vev. We thus introduce more suitable variables
\begin{equation}
	\sigma \coloneqq \varphi^{N}-\langle\varphi\rangle, \quad 
	\pi^{a} \coloneqq \varphi^{a} \quad(1\leq a<N)\,.
\end{equation}

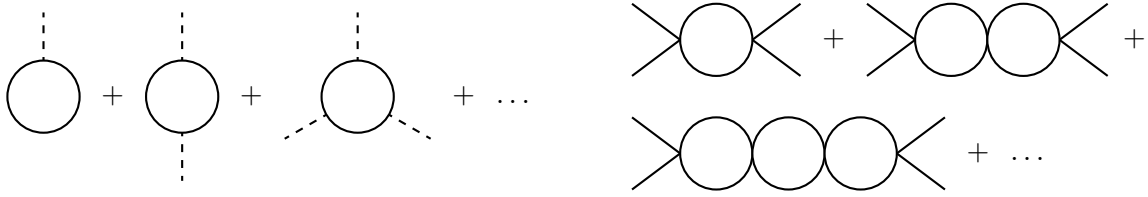
\begin{figure}
\begin{minipage}{.475\textwidth}
\pgfmathsetmacro{\bubbleSize}{0.6}
\pgfmathsetmacro{\tailSize}{1.4}
\begin{align*}
    \begin{tikzpicture}[yscale=0.8, xscale=0.8, baseline=-0.5ex, thick]
        \def\maxN{1}
        \foreach \i in {1,...,\maxN}
        {
        \draw[thick, dashed] (0,0) -- ({\tailSize*cos(360*\i/\maxN+90)},{\tailSize*sin(360*\i/\maxN+90))});
        }
        \filldraw[fill=white, thick] (0,0) circle (\bubbleSize);
    \end{tikzpicture}
    \,\,\,+\,\,\,
    \begin{tikzpicture}[yscale=0.8, xscale=0.8, baseline=-0.5ex, thick]
        \def\maxN{2}
        \foreach \i in {1,...,\maxN}
        {
        \draw[thick, dashed] (0,0) -- ({\tailSize*cos(360*\i/\maxN+90)},{\tailSize*sin(360*\i/\maxN+90))});
        }     
        \filldraw[fill=white, thick] (0,0) circle (\bubbleSize);
    \end{tikzpicture}
    \,\,\,+\,\,\,
    \begin{tikzpicture}[yscale=0.8, xscale=0.8, baseline=-0.5ex, thick]
        \def\maxN{3}   
        \foreach \i in {1,...,\maxN}
        {
        \draw[thick, dashed] (0,0) -- ({\tailSize*cos(360*\i/\maxN+90)},{\tailSize*sin(360*\i/\maxN+90))});
        }        
        \filldraw[fill=white, thick] (0,0) circle (\bubbleSize);
    \end{tikzpicture}
    \,\,\,+\,\,\,\dots
\end{align*}
\end{minipage}%
\hspace{0.05\textwidth}
\begin{minipage}{.475\textwidth}
\pgfmathsetmacro{\bubbleSize}{0.6}
\pgfmathsetmacro{\tailSize}{1.4}
\begingroup
\addtolength{\jot}{1em}
\begin{align*}
    &\begin{tikzpicture}[yscale=0.8, xscale=0.8, baseline=-0.5ex, thick]
        \def\maxN{1}
        \draw[thick] (2*\bubbleSize-\tailSize,\bubbleSize) -- (1*\bubbleSize,0);
        \draw[thick] (2*\maxN*\bubbleSize+\bubbleSize,0) -- (2*\maxN*\bubbleSize+\tailSize,\bubbleSize);
        \draw[thick] (2*\bubbleSize-\tailSize,-\bubbleSize) -- (1*\bubbleSize,0);
        \draw[thick] (2*\maxN*\bubbleSize+\bubbleSize,0) -- (2*\maxN*\bubbleSize+\tailSize,-\bubbleSize);
        \foreach \i in {1,...,\maxN}
        {
        \filldraw[fill=white, thick] (2*\i*\bubbleSize,0) circle (\bubbleSize);
        }
    \end{tikzpicture}
\,\,\,+\,\,\,
    \begin{tikzpicture}[yscale=0.8, xscale=0.8, baseline=-0.5ex, thick]
        \def\maxN{2}
        \draw[thick] (2*\bubbleSize-\tailSize,\bubbleSize) -- (1*\bubbleSize,0);
        \draw[thick] (2*\maxN*\bubbleSize+\bubbleSize,0) -- (2*\maxN*\bubbleSize+\tailSize,\bubbleSize);
        \draw[thick] (2*\bubbleSize-\tailSize,-\bubbleSize) -- (1*\bubbleSize,0);
        \draw[thick] (2*\maxN*\bubbleSize+\bubbleSize,0) -- (2*\maxN*\bubbleSize+\tailSize,-\bubbleSize);
        \foreach \i in {1,...,\maxN}
        {
        \filldraw[fill=white, thick] (2*\i*\bubbleSize,0) circle (\bubbleSize);
        }
    \end{tikzpicture}
\,\,\,+\\
    &\begin{tikzpicture}[yscale=0.8, xscale=0.8, baseline=-0.5ex, thick]
        \def\maxN{3}
        \draw[thick] (2*\bubbleSize-\tailSize,\bubbleSize) -- (1*\bubbleSize,0);
        \draw[thick] (2*\maxN*\bubbleSize+\bubbleSize,0) -- (2*\maxN*\bubbleSize+\tailSize,\bubbleSize);
        \draw[thick] (2*\bubbleSize-\tailSize,-\bubbleSize) -- (1*\bubbleSize,0);
        \draw[thick] (2*\maxN*\bubbleSize+\bubbleSize,0) -- (2*\maxN*\bubbleSize+\tailSize,-\bubbleSize);
        \foreach \i in {1,...,\maxN}
        {
        \filldraw[fill=white, thick] (2*\i*\bubbleSize,0) circle (\bubbleSize);
        }
    \end{tikzpicture}
\,\,\,+\,\,\,\dots
\end{align*}
\endgroup
\end{minipage}%
\caption{Solid lines denote pions $\pi^a$ and dashed lines denote the Hubbard-Stratonovich field $\chi$. Left: 1PI diagrams that contribute to the effective action at leading order in $1/N$. Right: Bubble diagrams that contribute to the $\pi\pi \to \pi \pi$ amplitude at leading order.}\label{fig:leadingOrderN}
\end{figure}

Before switching to these new variables, it is helpful for the large $N$ part of the analysis to introduce a Hubbard-Stratonovich (HS) field $\chi$, rewriting the Lagrangian as
\begin{equation}
	\mathcal{L}=\frac{1}{2} \partial_{\mu} \varphi^{a} \partial^{\mu} \varphi^{a}+\frac{1}{2} \frac{N}{\lambda_{0}} \chi^{2}-\frac{1}{2} \chi \varphi^{a} \varphi^{a}-\frac{N \mu_{0}^{2}}{\lambda_{0}} \chi\,.
\end{equation}
The introduction of $\chi$ is a combinatorial trick which greatly simplifies the large-$N$ analysis: there is now only one three-point vertex $\chi\varphi^2$, all factors of $\frac{1}{N}$ come from the $\chi$ propagator, and all factors of $N$ come from $\varphi$ loops. Hence, to leading order in large $N$, the only diagrams that contribute to the effective action are those 1PI diagrams with a single loop of $\varphi$ and an arbitrary number of insertions of the field $\chi$, as shown in Figure \ref{fig:leadingOrderN}.

Putting both of the pieces together, the effective action is found to be \cite{Coleman:1974jh}
\begin{align}
		\Gamma=\int d^{n} \chi& \left[\frac{1}{2} \pi^{a} \Box \pi^{a}+\frac{1}{2} \sigma \Box \sigma+\frac{1}{2} \frac{N}{\lambda_{0}} \chi^{2}-\chi \sigma\langle\varphi\rangle-\frac{1}{2} \chi \sigma^{2}\nonumber\right.\\&\left.
		-\frac{1}{2} \chi \pi^{a} \pi^{a}
		-\left(\frac{N \mu_{0}^{2}}{\lambda}+\frac{1}{2}\langle\varphi\rangle^{2}\right) \chi\right]
		-\frac{1}{2} N \operatorname{tr}\ln \left(-\Box+\chi\right). \label{eq:effectiveaction}
\end{align}
The HS field $\chi$ is taken to have zero expection value $\langle \chi \rangle = 0$. This is guaranteed by requiring the sum of all tadpole diagrams to vanish, i.e. terms linear in $\chi$ in the expansion, which gives rise exactly to Equation \eqref{eq:vev}.

Following \cite{Coleman:1974jh}, we compute the propagators from the effective action in Equation \eqref{eq:effectiveaction}. The propagators for the pions take the expected form $D_{\pi^a\pi^b}  = \delta^{ab}/p^2$. On the other hand, there is a mixing between $\sigma$ and $\chi$, i.e.
\begin{equation}
	-\frac{1}{2} \begin{pmatrix}
		\sigma & \chi
	\end{pmatrix} D^{-1}(p^2) 
	\begin{pmatrix}
		\sigma\\
		\chi
	\end{pmatrix},
\end{equation}
with the non-diagonal matrix
\begin{equation}
	D^{-1}(p^2) = 
	    \begin{pmatrix}
		p^{2} & \langle\varphi\rangle \\
		\langle\varphi\rangle\,\, & -\frac{N}{\lambda}-\frac{N}{16 p}
	\end{pmatrix}.
\end{equation}
Inverting this matrix, we find the propagators are double-valued functions of $p^2$
\begin{align}
	D_{\sigma\sigma} &= \frac{p+\frac{\lambda}{16}}{p\left( p^2 + \frac{\lambda}{16}p-2\mu^2\right)}\\
	D_{\chi\chi} & = -\frac{\lambda}{N} \frac{p^2}{p^2 + \frac{\lambda}{16} p -2\mu^2}\label{eq:propagatorchichi}\\
	D_{\sigma\chi} = D_{\chi\sigma} & = \sqrt{\frac{-2\mu^2\lambda}{N}} \frac{1}{p^2 + \frac{\lambda}{16}p-2\mu^2}\,,
\end{align}
where $p$ is the positive square root of $p^2$. Importantly, $D_{\sigma\sigma}$ diverges at $p=0$ because of a pole in $p$, and not because of a pole in $p^2$ (in $p^2$ this is a branch point from the cut associated with two-$\pi$ intermediate states). Both propagators have poles on the second sheet of the $p^2$-plane, given by the zeros of $\det D^{-1}(p^2)$
\begin{equation}
	p_{\pm} =  -\frac{\lambda}{32} \pm i \sqrt{-2 \mu^{2}-\left(\frac{\lambda}{32}\right)^{2}}\,.\label{eq:res}
\end{equation}
Physically, the two poles are interpreted as two resonances. All of the above appears in the analysis of \cite{Coleman:1974jh}.

For simplicity, we will assume that $p_{\pm}$ have both nonzero real and imaginary components, which motivates the definition
\begin{equation}
    \cos(\theta) \coloneqq \frac{\lambda}{16\sqrt{-8\mu^{2}}}<1\,.
\end{equation}
However, it will become clear that this is not a restrictive assumption in any way, since the resonance exists as long as $\lambda\neq 0$ and no further bound is required for our derivation.\footnote{If $\lambda$ is large enough $p_{\pm}$ will become purely real. This is however not an issue since one can check that the two poles are always on the unphysical sheet. Indeed, all of the discussion in this article holds for $\lambda>0$.} Moreover, recall that boundedness of the potential requires $\lambda>0$.

Now let us briefly discuss the continuation to Lorentzian signature, which will be necessary later. In order to see the correct continuation in the energy variable, consider the Riemann surface for the Lorentzian\footnote{From here out, when we use the variable $p$ in scattering amplitudes, we are working in Euclidean space; when we use variables like $s$, $t$, $u$, and $\omega$, we are working in Lorentzian.} momentum variable $s:=\omega^2$. To identify the correct Wick rotation, we locate the branch cut of the effective propagator along the positive real $s$ axis; this starts at $s=0$. From here, only the choice $p\to -i\omega$ is physical, sending the resonances in Equation \eqref{eq:res} to the second-sheet in the $s$-variable. 

In the small coupling limit, the zeroes of the Wick rotated propagator are at
\begin{equation}
    \omega \to\pm\sqrt{-2\mu^2}-i\frac{\lambda}{32}  \label{eq:wpmres}\,.
\end{equation}
From this, one can see how the resonances approach the tree level expression $\pm \sqrt{-2\mu^{2}}$ as $\lambda \to 0$, and how they emerge at higher-loop order because of the non-zero Higgs width $\lambda$. Now, at leading order in $\lambda$ and close to $\omega = +\sqrt{-2\mu^{2}}$
\begin{equation}
    \frac{1}{\omega^2+i\frac{\lambda}{16}\omega+2\mu^2} \to \frac{1}{\omega^{2} + i\frac{\lambda}{16}\sqrt{-2\mu^2}+2\mu^2}\,,
\end{equation}
which is of the standard Breit-Wigner form $\frac{1}{s-m^2 +im\Gamma}$. Clearly since this is valid only close to $\omega = + \sqrt{-2\mu^{2}}$, only the root $\omega_{+} = +\sqrt{-2\mu^{2}} - i \epsilon$ of the denominator is meaningful. This also has the correct sign for the $i\epsilon$-prescription for tree level propagators where the identification $\epsilon = \sqrt{-2\mu^2}\frac{\lambda}{16}$ has been made (the analogous analysis holds close to $\omega=-\sqrt{-2\mu^{2}}$). Notice that if the wrong continuation $p\to i\omega$ had been used, then the tree level result for the propagators would have appeared with the wrong sign for the $i\epsilon$-prescription.

\subsection{Three and Four-point Momentum Amplitudes}\label{sec:threeFourAmps}
We are now in position to discuss scattering amplitudes. For example, it is easy to write down the $\pi^a\pi^b\rightarrow \pi^c\pi^d$ amplitude using the full propagator $D_{\chi\chi}$ in \eqref{eq:propagatorchichi} repeatedly
\begin{align}
    T_{4}^{ab,cd}(p_{1},p_{2},p_{3},p_{4}) = 
        -\frac{\lambda}{N} \bigg[&\frac{\,(p_{12})^{2}\,\delta^{ab} \delta^{cd}}{(p_{12})^{2} + \sqrt{(p_{12})^{2}} \lambda/16 - 2\mu^{2}} +  \frac{\,(p_{13})^{2}\,\delta^{ac} \delta^{bd}}{(p_{13})^{2} + \sqrt{(p_{13})^{2}} \lambda/16 - 2\mu^{2}} \nonumber \\[0.3cm] &\hspace{3.5cm}+ \frac{\,(p_{14})^{2}\,\delta^{ad} \delta^{bc}}{(p_{14})^{2} + \sqrt{(p_{14})^{2}} \lambda/16 - 2\mu^{2}} \bigg]\,, \label{eq:Tpipipipi}
\end{align}
where $(p_{ij})^{2} = (p_{i} + p_{j})^{2}$ and we have used the convention that all the momenta are incoming.

The tree-level result, obtained as $\lambda\to 0$, corresponds to the scalar tree-level amplitude studied by Lam and Shao in \cite{lamShao}. Indeed, as $\lambda$ stays positive but becomes small, the two poles approach the point $p^2 = 2\mu^2$ in the physical region with the physical $i\epsilon$ deformation. However, an important difference between \eqref{eq:Tpipipipi} and the amplitude of \cite{lamShao} is the presence of additional powers of $(p_{ij})^{2}$ in the numerators of \eqref{eq:Tpipipipi}. This reflects the Adler's zero phenomenon: the fact that the scattering of Goldstone bosons must vanish as $(p_{ij})^{2} \rightarrow 0$ \cite{Adler:1964um}. As we will see in Section \ref{sec:ONAmplitude}, this soft behaviour will be reflected in the analyticity domain for the celestial amplitude as a function of the boost weight $\Delta$. Note further that the full large $N$ amplitude is finite in the high-energy limit at fixed scattering angle. Such UV softness will be also essential for the existence of a convergence domain in $\Delta$.

The $\pi^a\pi^b \rightarrow \sigma$ scattering amplitude can be found by applying the standard Lehmann Symanzik Zimmermann (LSZ) reduction formula to the three-point function $\expval{\pi^{a}\left(p_{1}\right) \pi^{b}\left(p_{2}\right) \sigma\left(p_{3}\right)}$. Recall that the LSZ reduction formula works by bringing the field operator to the mass shell and studying the poles of the correlation functions. The $S$ matrix elements are simply given by the corresponding residues after we take care of the normalization coming from wavefunction renormalization. 

More explicitly, by Wick contraction, we find the three point function to be
\begin{align}
	\left\langle\pi^{a}\left(p_{1}\right) \pi^{b}\left(p_{2}\right) \sigma\left(p_{3}\right)\right\rangle &= D_{\pi^a\pi^c}(p_1) D_{\pi^b\pi^c}(p_2) D_{\sigma\chi}(p_3)\,, \nonumber \\
	& = \delta^{a b} \frac{1}{p_{1}^{2}} \frac{1}{p_{2}^{2}} \sqrt{\frac{-2 \mu^{2} \lambda}{N}} \frac{1}{p_{3}^{2}+\frac{\lambda}{16} p_{3}-2 \mu^{2}}\,.
\end{align}
If we bring all three external particles on shell, we can read off S matrix elements as
\begin{align}
	\left\langle\pi^{a}\left(p_{1}\right) \pi^{b}\left(p_{2}\right) \sigma\left(p_{3}\right)\right\rangle \sim \frac{1}{p_{1}^{2}} \frac{1}{p_{2}^{2}} \frac{\sqrt{Z}}{p_3^2-p_{\pm}^2} \langle \sigma(p_{\pm})|\pi^a(p_1)\pi^b(p_2)\rangle\,,
\end{align}
where $Z_{\pm}$ denotes the wavefunction renormalization appearing in the residue at the poles $D_{\sigma\sigma} \sim Z_{\pm}/(p^2-p^2_\pm)$. Altogether we have
\begin{equation} \label{eq:3point}
	T_3^{ab}(p_{\pm}) \coloneqq \langle \sigma(p_\pm)|\pi^a(p_1)\pi^b(p_2)\rangle = \sqrt{\frac{-\mu^2 \lambda}{N}} \frac{-2p_\pm}{\sqrt{p_\mp(p_\mp-p_\pm)}} \delta^{ab}\,.
\end{equation}

Finally, as a consistency check, notice that
\begin{equation} \label{fourpointfromthreepointsquare}
    \underset{p_{12}^{2} = p^2_{\pm}}{\mathbf{Res}} \langle \pi^c(p_3) \pi^d(p_4) | \pi^a(p_1) \pi^b(p_2) \rangle =  \langle\pi^c(p_3)\pi^d(p_4) |  \sigma(p_\pm)\rangle  \langle \sigma(p_\pm)|\pi^a(p_1)\pi^b(p_2)\rangle,
\end{equation}
where only the first term in \eqref{eq:Tpipipipi} contributes.

\section{The Celestial \texorpdfstring{$O(N)$}{O(N)} Amplitude}\label{sec:CelestialON}
In this section, we introduce the celestial amplitude for massless external particles, as defined in \cite{stromingerCelestial, pasterskiShao}, and then apply it to $2\to2$ pion scattering in the symmetry broken phase of the 3d $O(N)$ model. We see how the all-loop resummation affects convergence of the Mellin integral, study the EFT expansion, and examine the constraints of scattering bosons for the celestial four-point function.

\subsection{Massless Scattering and the Celestial Circle}\label{sec:celestialSetup}
The celestial sphere formalism recasts the S-matrix elements of a Poincar\'e-invariant theory in $\bbR^{1,d+1}$ as ``correlation functions'' on the $d$-dimensional celestial sphere by re-writing the scattering amplitude in a basis of $SO(1,d+1)$ covariant wavefunctions, rather than the usual plane-wave basis given by the eigenfunctions of translation operators \cite{stromingerCelestial, pasterskiShao} (see also \cite{deBoer:2003vf, fullVirasoro:3} for related work). Since we will only be studying massless $2\to 2$ scattering, we will focus on just the details we need for this case.

Given a massless particle in $\bbR^{1,d+1}$ its momentum can be parametrized as
\begin{equation}
    p^\mu= \epsilon \omega (1+\abs{\vec{x}}^2, 2\vec{x}, 1-\abs{\vec{x}}^2)\,,
\end{equation}
where $\epsilon = \pm 1$ designates whether it is outgoing/incoming, $\omega > 0$ is a positive energy scale, and $\vec{x} \in \bbR^d$ gives a position on the (past or future) celestial sphere: the projective null-cone in momentum space. Following \cite{Strominger:2017zoo}, we identify past ($\epsilon=-1$) and future ($\epsilon=+1$) spheres via antipodal matching.

For $n$ massless external particles in $\bbR^{1,d+1}$, the \textit{celestial amplitude} is obtained from the flat space scattering amplitude $T(s,t)$ by a Mellin transform in each of the scale variables
\begin{equation}
    \widetilde{\mathcal{A}}(\Delta_i,x_i) 
        = \prod_{i=1}^{n} \int_{0}^{\infty} d \omega_i\, \omega_i^{\Delta_i-1}\, T(s,t) \, \delta^{(d+2)}\left(\textstyle{\sum}\, p_i\right)\,. \label{eq:celestialAmplitude}
\end{equation}
The celestial amplitude's construction makes its $SO(1,d+1)$ covariance manifest and ensures that it is at least as constrained as an $n$-point correlator in a ``typical'' $d$-dimensional CFT \cite{stromingerCelestial, pasterskiShao, lamShao}. Readers familiar with the embedding space formalism used in, for example, AdS/CFT or the conformal bootstrap, may find this procedure familiar \cite{Dirac:1936fq, Mack:1969rr, Weinberg:2010fx, Costa:2011mg, Costa:2011dw, simmonsDuffin:shadows}.

However, the original amplitude transformed under the full Poincar\'e group, which leads to additional constraints on the celestial amplitudes beyond the $SO(1,d+1)$ covariance of a generic CFT correlator. As highlighted in \cite{celestialConstraints, Law:2020tsg, Law:2020xcf, Arkani-Hamed:2020gyp, Atanasov:2021cje}, the invariance under translations leads to phenomena like distribution valued conformal correlators if $n \leq d+2$, and special restrictions for the celestial amplitude on the external scaling dimensions $\Delta_i$ (see \cite{celestialConstraints} for specifics).

For massless $2\to 2$ amplitudes in, say, $d=1$ and $d=2$, a straightforward evaluation of equation \eqref{eq:celestialAmplitude} demonstrates a factorization into a universal ``kinematic part'' $X(\Delta_i,x_i)$, and a ``dynamical part'' $\calA(\Delta_T,z)$ which depends on the scattering amplitude $T(s,t)$, i.e.
\begin{align}
    \widetilde{\mathcal{A}}_{d=1}(\Delta_i,x_i) 
        &= X_{d=1}(\Delta_i,x_i) \calA(\Delta_T,z)\,,\label{eq:3d4ptFunc}\\
    \widetilde{\mathcal{A}}_{d=2}(\Delta_i,x_i) 
        &= X_{d=2}(\Delta_i,x_i) \calA(\Delta_T,z)\,.
\end{align}
Here $\Delta_T = \sum \Delta_i$ and $z$ is the conformal cross-ratio
\begin{equation}
    z \coloneqq \frac{x_{12}x_{34}}{x_{13}x_{24}}\,.
\end{equation}

To see this, note that on the support of the momentum conserving delta function in Equation (\ref{eq:celestialAmplitude}) the Mandelstam variables can be written as
\begin{equation}
    s = -4\frac{x_{12}^2x_{13}x_{14}}{x_{23}x_{24}}\omega_1^2\,,\quad
    t = -\frac{s}{z}\,,\quad
    u = \frac{1-z}{z}s\,.
\end{equation}
Moreover, for each of the three physical scattering channels, we have a different admissible range of cross-ratio $z$
\begin{align}
    \text{$u$-channel: } & z\in(-\infty,0)\,,\\
    \text{$t$-channel: } & z\in(0,1)\,,\\
    \text{$s$-channel: } & z\in(1,\infty)\,.
\end{align}
See \cite{lamShao, Chang:2021wvv, Mizera:2022sln} for in-depth discussions.

Evaluating three of the delta functions, the kinematic parts can be written explicitly as
\begin{align}
    X_{d=1}(\Delta_i,x_i)
        &= \left(\prod_{i < j} x_{ij}^{\Delta_T/3-\Delta_i-\Delta_j}\right) 2^{-\Delta_T+1} \left(\frac{z}{\sqrt{1-z}}\right)^{1-\Delta_T/3}\,,\\
    X_{d=2}(\Delta_i,x_i)
        &= \left(\prod_{i < j} x_{ij}^{h/3-h_i-h_j}\bar{x}_{ij}^{\bar{h}/3-\bar{h}_i-\bar{h}_j}\right)\delta(z-\bar{z})2^{-\Delta_T+2}\abs{\frac{z}{\sqrt{1-z}}}^{-\frac{\Delta_T}{3}}\,.
\end{align}
Note that the 1d kinematic piece does not possess the infamous momentum conservation $\delta(z-\bar{z})$ which shows up for massless $2\to2$ scattering in 2d.

Meanwhile, the dynamical data is contained in $\calA(\Delta_T,z)$, which boils down to a single Mellin transform\footnote{For convenience, we will write $T(\omega,z)$ whenever we refer to a general amplitude in the $(\omega,z)$ variables i.e. before Mellin-transforming, as no confusions should arise.} of $T(\omega,z)$ in center of mass energy $\omega \coloneqq \sqrt{\epsilon_s s}$ where $\epsilon_s \coloneqq \epsilon_1 \epsilon_2$
\begin{equation}
    \calA(\Delta_T,z) = \int_0^\infty d\omega\,\omega^{\Delta_T-4}\,T \Big( \epsilon_s\omega^2,-\epsilon_s\frac{\omega^2}{z} \Big) \,.\label{eq:MellinInEnergy}
\end{equation}

Altogether, any  celestial amplitude $\widetilde{\mathcal{A}}_{d=1}(\Delta_i,x_i)$ takes the form of a 1d CFT correlator
\begin{equation}
    \widetilde{\mathcal{A}}_{d=1}(\Delta_i,x_i)
        =\frac{\left(\frac{x_{14}^2}{x_{24}^2}\right)^{-\frac{1}{2}\Delta_{12}}\left(\frac{x_{14}^2}{x_{13}^2}\right)^{\frac{1}{2}\Delta_{34}}}{\left(x_{12}^2\right)^{\frac{1}{2}(\Delta_1+\Delta_2)}\left(x_{34}^2\right)^{\frac{1}{2}(\Delta_3+\Delta_4)}} f(z)
\end{equation}
with the conformally invariant part (aka ``stripped four-point function'')
\begin{equation}
    f(z) = 2^{-\Delta_T+1} z(1-z)^{-\frac{1}{2}(1-\Delta_{12}+\Delta_{34})} \calA(\Delta_T,z)\,.
\end{equation}
For four identical particles with $\Delta_i = \Delta$, this reduces to
\begin{align}
    \widetilde{\mathcal{A}}\left(\Delta, x_{i}\right)
        &=\frac{1}{\left|x_{12}\right|^{2 \Delta}\left|x_{34}\right|^{2 \Delta}} f(z)\,,\\[0.2cm]
    f(z)
        &=2^{-\Delta_T+1} \frac{|z|}{\sqrt{|z-1|}} \calA(\Delta_T, z)\,.
\end{align}

\subsection{The \texorpdfstring{$O(N)$}{O(N)} Amplitude} \label{sec:ONAmplitude}
We may now compute and analyze the $O(N)$ celestial amplitude by directly applying the tools from the previous section. We focus on the physical $s$-channel $12\to34$ so that $z\in(1,\infty)$ for concreteness and compare each of our results to the (2+1)d massless scalar scattering studied in \cite{lamShao}. As clear from the previous subsection, to perform the Mellin transform \eqref{eq:MellinInEnergy} we need to recast the amplitude \eqref{eq:Tpipipipi} in Lorentzian signature. The physical analytic continuation consistent with the $i \epsilon$-prescription at small $\lambda$ leads to
\begin{align} 
T_{4}^{ab,cd}\Big(\omega^{2}, - \frac{\omega^{2}}{z} \Big)
        = - \frac{\lambda}{N} \bigg[\frac{\omega^{2}\delta^{ab}\delta^{cd}}{\omega^{2} + i\frac{\lambda}{16} \omega +2\mu^{2}}
        &+ \frac{\omega^{2} \delta^{ac}\delta^{bd}}{\omega^{2} +  \frac{\lambda}{16}\sqrt{z}\omega - 2 \mu^{2} z }
        \nonumber \\*
        &+ \frac{\omega^{2}\delta^{ad}\delta^{bc}}{\omega^{2} + \frac{\lambda}{16} \sqrt{\frac{z}{z-1}}\omega -2\mu^{2}\frac{z}{z-1}}\bigg]\,. \label{4pt-omegaz}
\end{align}
For future reference, let us spell out the position of the different poles in this expression:
\begin{align}
    \omega^{\mp}_{ab,cd} & =\frac{1}{2}\Bigg(-i\frac{\lambda}{16} \pm \sqrt{-\Big( \frac{\lambda}{16} \Big)^{2} - 8\mu^{2}}\Bigg), \label{omegaabcd}\\[0.15cm] \omega^{\mp}_{ac,bd} & = \frac{1}{2}\sqrt{z} \Bigg( -\frac{\lambda}{16} \pm i \sqrt{-\Big(\frac{\lambda}{16}\Big)^{2} - 8\mu^{2}} \Bigg),  \label{omegaacbd}\\[0.15cm] \omega^{\mp}_{ad,bc} & = \frac{1}{2}\sqrt{\frac{z}{z-1}} \Bigg( -\frac{\lambda}{16} \pm i \sqrt{-\Big(\frac{\lambda}{16}\Big)^{2} - 8\mu^{2}} \Bigg), \label{omegaadbc}
\end{align}
where $\omega^{\mp}_{ab,cd}$ stands for the poles associated to the term proportional to $\delta^{ab}\delta^{cd}$ in \eqref{4pt-omegaz},\footnote{The ``mismatch'' between the signs in the left and right sides of these expressions come from the fact that under Wick rotation, $p_{\pm}$ in \eqref{eq:res} becomes $\omega_{\pm}$.} and similarly for $\omega^{\mp}_{ac,bd}$ and $\omega^{\mp}_{ad,bc}$.

All three of the contributions in \eqref{4pt-omegaz} require Mellin transforms of the form\footnote{See entry (12) of Section 6.2 of \cite{bateman1954tables}.}
\begin{equation} \label{batemanformula}
    \int_{0}^{\infty} \frac{\omega^{\Delta_T-2}}{\omega^2+ 2a\cos\theta\,\omega + a^2} d\omega 
        = \pi a^{\Delta_T-3} \frac{\sin{[(\Delta_T-2)\theta]}}{\sin{\theta} \sin{(\pi \Delta_T)}}\,,
\end{equation}
where, to avoid IR and UV divergences, the integral requires $-1 < \Re(\Delta_T) -2 < 1$. We also note that none of the three terms have poles on the positive real axis (recall $\mu^2 < 0$ and $z\in(1,\infty)$), hence the function is integrable and admits a Mellin transform in the usual sense. This is in contrast with the tree-level case $\lambda\to 0$ where the poles pinch the integration region and need to be regulated \cite{lamShao}. In our case the finite coupling acts as a natural regulator.

Putting this together, the Mellin-transformed $s$-channel amplitude takes the following form:
\begin{equation} \label{Mtransform}
    \calA^{ab,cd}(\alpha,z) = 2^{2+2\alpha}\calN \calN_{\alpha} \left[ \delta^{ab}\delta^{cd}e^{i \pi \alpha} + \delta^{ac}\delta^{bd}z^{\alpha} + \delta^{ad}\delta^{bc}\left( \frac{z}{z-1} \right)^{\alpha} \right], \quad z\in(1,\infty),   
\end{equation}
where we have defined 
\begin{equation}
    \calN \coloneqq \frac{\pi \lambda}{N \sin\theta}\,,
        \qquad \calN_{\alpha} \coloneqq \frac{(-2\mu^2)^\alpha}{2^{2+2\alpha}} \frac{\sin[(2\alpha+1)\theta]}{\sin(2\pi\alpha)}\,,
        \qquad \alpha \coloneqq (\Delta_{T}-3)/2\,.
\end{equation}
For the case of four identical external weights $\Delta_i = \Delta$ the (conformally invariant piece of the) celestial amplitude is
\begin{align}
    f^{ab,cd}_{s}(z) 
        & = \frac{1}{2^{2+2\alpha}} \frac{z}{\sqrt{z-1}}\calA(\alpha,z),\\
        & = \calN \calN_{\alpha} \frac{z}{\sqrt{z-1}} \left[ \delta^{ab}\delta^{cd}e^{i \pi \alpha} + \delta^{ac}\delta^{bd}z^{\alpha} + \delta^{ad}\delta^{bc} \left( \frac{z}{z-1} \right)^{\alpha} \right],\quad z\in(1,\infty). \label{celestialamplitude}
\end{align}
We find that (up to color structures) this result has the same $z$-dependence as the tree-level example originally studied by Lam and Shao in \cite{lamShao}, which is itself closely related to a Generalized Free Field theory, with an overall kinematic pre-factor and an imaginary piece which marks the physical channel. 

It is instructive to see how the result of \cite{lamShao} appears in the tree-level approximation. We approach this limit by taking $\lambda\to 0$ while keeping $-\mu^{2}$ fixed, which amounts to take $\theta\to \pi/2$. The result is
\begin{equation} \label{Lam-Shao-recovered}
   f^{ab,cd}_{s}(z) \xrightarrow[\text{$\lambda\to 0$}]{\text{}} \frac{\pi \lambda  \big( -2\mu^{2} \big)^{\alpha}}{2^{ 4 \Delta } N \cos{(2 \pi\Delta)}} \frac{ z}{\sqrt{z-1}} \left[ \delta^{ab}\delta^{cd}e^{i \pi \alpha} + \delta^{ac}\delta^{bd}z^{\alpha} + \delta^{ad}\delta^{bc} \left( \frac{z}{z-1} \right)^{\alpha} \right],
\end{equation}
which is precisely the result of \cite{lamShao} up to an extra power of $-2\mu^2=M^2$. This is nothing but the effect of the extra power of $\omega^2$ dictated by Adler zero.

One distinct effect of the all-loop resummation in comparison with the tree-level amplitude, besides the modification of the prefactors $\calN$ and $\calN_{\alpha}$, is the modification in the convergence region for the Mellin integral. This modification occurs because of an overall extra power of $\omega^2$ in the numerator of the resummed $T_4^{ab,cd}$ relative to the tree-level result. Indeed, we have that:
\begin{align}
    \text{Tree-Level:}\quad  & 3 < \Re(\Delta_T) < 5,\\
    \text{Resummed:}\quad    & 1 < \Re(\Delta_T) < 3.
\end{align}
Interestingly, this convergence range for the resummed result allows all of the scattering $\Delta_i$ to be on the principal continuous series of $SO(1,2)$, i.e. $\Delta_i = \frac{1}{2}+i\ell$, $\ell\in\bbR$. Famously, conformal primary wavefunctions with $\Delta$ on the principal continuous series form a complete set of normalizable solutions to the massless Klein-Gordon equation, with respect to the Klein-Gordon inner product \cite{pasterskiShao}.

\subsection{Analyticity in $\Delta$ and EFT Expansion}\label{sec:EFTExpansion}

Let us close this subsection making explicit the analytic structure of the celestial amplitude in the complex $\Delta_T$-plane. As expected from a field theory result, we observe that there are infinitely many simple poles at both positive and negative values of $\Delta_{T}$ (in contrast to a quantum gravity/string theory result, as discussed in \cite{Arkani-Hamed:2020gyp}). Clearly there are simple poles at integers $\Delta_T = m$ for $m \leq 1$ and $m \geq 3$. At $m=2$, the celestial amplitude is finite in accordance with our aforementioned comment on the amplitude being well-defined on the principal continuous series $\Delta_i = \frac{1}{2}+i\ell$, $\ell\in\bbR$. 

As pointed out in \cite{Arkani-Hamed:2020gyp}, the residues of the celestial amplitude in $\Delta_{T}$ are not arbitrary but are in direct correspondence to EFT operators at four points. They correspond to the coefficients of the low energy expansion of the corresponding momentum space amplitude $T(\omega,z)$ if the poles in $\Delta_{T}$ organize in the left side of the $\Delta_{T}$-plane, and to the coefficients of the high energy expansion if the poles organize in the right side of the $\Delta_{T}$-plane. In our context, these are the $m \leq 1$ and $m \geq 3$ families of poles respectively. 

Let us then illustrate a specific all-loop resummed instance of this observation in both the UV and IR. This means we perform the expansion of the field theory result as $\omega \to 0$:\footnote{This expansion can be performed with the aid of the series
\begin{equation*}
    \sum_{k=1}^{\infty} p^{k}\sin{(kx)} = \frac{p \sin{x}}{1-2p\cos{x}+p^{2}}\,,  \qquad |p| < 1\,.
\end{equation*}}
\begin{align} 
    T(\omega,z) 
        = -\frac{\lambda}{N} \sum_{k=1}^{\infty} \Bigg[ \delta^{ab}\delta^{cd} \Bigg( \frac{e^{i\frac{\pi}{2}} \omega}{\sqrt{-2\mu^{2}}} \Bigg)^{k+1}
        &+ \delta^{ac}\delta^{bd} \Bigg( \frac{e^{i\pi} \omega}{\sqrt{-2\mu^{2}}\sqrt{z}} \Bigg)^{k+1} \nonumber\\
        &+ \delta^{ad}\delta^{bc}\Bigg( \frac{e^{i\pi} \omega \sqrt{\frac{z-1}{z}}}{\sqrt{-2\mu^{2}}} \Bigg)^{k+1}\Bigg] \frac{\sin{(k \theta)}}{\sin{\theta}} \,, \label{lowenergyexpansion}
\end{align}
whose coefficients indeed reproduce the residues of \eqref{Mtransform} as $\Delta_{T} \to m$ for $m \leq 1$. Similarly, the high energy expansion
\begin{align}
    T(\omega,z) 
        = -\frac{\lambda}{N} \sum_{k=1}^{\infty} \Bigg[ \delta^{ab}\delta^{cd} \Bigg(e^{-i\frac{\pi}{2}} \frac{ \sqrt{-2\mu^{2}}}{\omega} \Bigg)^{k-1}
        &+ \delta^{ac}\delta^{bd} \Bigg( \frac{e^{i\pi} \sqrt{-2\mu^{2}}\sqrt{z}}{\omega} \Bigg)^{k-1}\nonumber\\
        &+ \delta^{ad}\delta^{bc} \Bigg( \frac{e^{i\pi} \sqrt{-2\mu^{2}} \sqrt{\frac{z}{z-1}}}{\omega} \Bigg)^{k-1}\Bigg] \frac{\sin{(k \theta)}}{\sin{\theta}} \,, \label{highenergyexpansion}
\end{align}
is analogously related to the poles of \eqref{Mtransform} as $\Delta_{T} \to m$ for $m \geq 3$.

\subsection{S-Matrix Permutation Symmetry}\label{sec:SMatCrossing} 
Repeating the results of the previous section for different physical regions, we can write the three celestial amplitudes as:
\begin{alignat}{3}
    f_u^{abcd}(z) &= 
        \calN \calN_{\alpha} \frac{-z}{\sqrt{1-z}} \left(\delta^{ab}\delta^{cd}+\delta^{ac}\delta^{bd}(-z)^{\alpha}+\delta^{ad}\delta^{bc}e^{i\pi\alpha}\left(\frac{-z}{1-z}\right)^\alpha\right)\,, 
        && \quad z\in (-\infty,0)\,,\label{eq:CelestialONfu}\\
    f_t^{abcd}(z) &= 
        \calN \calN_{\alpha} \frac{z}{\sqrt{1-z}} \left(\delta^{ab}\delta^{cd}+\delta^{ac}\delta^{bd}e^{i\pi\alpha}z^{\alpha}+\delta^{ad}\delta^{bc}\left(\frac{z}{1-z}\right)^\alpha\right)\,, 
        && \quad z\in (0,1)\,,\label{eq:CelestialONft}\\
    f_s^{abcd}(z) &= 
        \calN \calN_{\alpha} \frac{z}{\sqrt{z-1}} \left(\delta^{ab}\delta^{cd}e^{i\pi\alpha}+\delta^{ac}\delta^{bd}z^{\alpha}+\delta^{ad}\delta^{bc}\left(\frac{z}{z-1}\right)^\alpha\right)\,, 
        && \quad z\in (1,\infty)\,.\label{eq:CelestialONfs}
\end{alignat}
The functions are not analytic continuations of one another in $z$, but ``permutation symmetry'' (i.e. relabelling the external scattering states) of the scattering amplitude implies:
\begin{alignat}{3}
    f_t^{abcd}\left(\frac{z}{z-1}\right) 
        &= f_u^{bacd}(z) = f_u^{abdc}(z)\,, 
        && \quad z\in(-\infty,0)\,,\\
    z^{2\Delta} f_t^{abcd}\left(\frac{1}{z}\right) 
        &= f_s^{acbd}(z) = f_s^{dbca}(z)\,, 
        && \quad z\in(1,\infty)\,,
\end{alignat}
as well as
\begin{equation}
    z^{-2\Delta}f_t^{abcd}\left(\frac{1}{z}\right)
        = (1-z)^{-2\Delta} f_t^{adcb}(z) 
        = (1-z)^{-2\Delta} f_t^{cbad}(z)\,, 
         \quad z\in(0,1)\,.\\
\end{equation}
These relations are the color analogs of those given for the 1d CCFT coming from massless scattering in \cite{lamShao}. 

These equations are all familiar from 1d CFT as we explain in Appendix \ref{sec:1dCFTGeneral} (see also Section 2 of \cite{Mazac:2018qmi}). In short, a 1d CFT correlator is defined piecewise on the three disconnected regions $(-\infty,0)$, $(0,1)$, and $(1,\infty)$. Similar to our 1d celestial amplitudes, the pieces cannot generally be related by analytic continuation in $z$, but if it is a correlator of bosons, then the pieces are related by Bose symmetry. Those relations implied on the 1d boson correlators are precisely those we have found above.

In other words, the original Bose symmetry of the S-matrix implies relationships between the celestial amplitudes in their different physical regions, and those relations are necessary for the celestial amplitude to literally ``be a CFT four point function of bosons'' on the celestial circle. This is consistent with the extrapolate dictionary \cite{Pasterski:2021dqe,Donnay:2022sdg}.

Note: this is not crossing symmetry, which is a statement about obtaining scattering amplitudes in different physical channels by analytic continuation of the complexified S-matrix. First steps to understanding crossing symmetry in celestial holography have appeared in \cite{Mizera:2022sln} (see also \cite{Hannesdottir:2022bmo} for related discussion).

\section{The Imaginary Part of the Celestial Amplitude and Resonances}\label{sec:GenOpticalThm}
In this section we study the analytic properties of scattering amplitudes in celestial variables, extending the work of \cite{Chang:2021wvv} to the non-perturbative regime by allowing for unstable resonances (like those in the $O(N)$ model). In particular, we see how poles associated with unstable particles control the imaginary part of the correlator relevant to the optical theorem. We will check all of our results explicitly in the $O(N)$ model case, whose resonances we have reviewed in Section \ref{sec:effDescON}.

\begin{figure}[t]
	\centering
	\begin{tikzpicture}[]
        \def\bigradius{3}
        \def\gap{3/10}
        
        \draw[help lines,<->] (-1*\bigradius, 0) -- (1*\bigradius,0);
        \draw[help lines,<->] (0, -1*\bigradius) -- (0, 1*\bigradius);
        
        \draw[-,decorate,decoration={snake,amplitude=.4mm,segment length=2mm,post length=1mm},red] (1*\bigradius,-1*\gap) -- (0,0) node[midway, above] {};
        
        \draw[-,decorate,decoration={snake,amplitude=.4mm,segment length=2mm,post length=1mm},black!50!green] (+1*\gap,+1*\bigradius) -- (0,+4*\gap) node[midway, above] {};
        \draw[-,decorate,decoration={snake,amplitude=.4mm,segment length=2mm,post length=1mm},black!50!green] (-1*\gap,-1*\bigradius) -- (0,-4*\gap) node[midway, above] {};
        
        \draw [->- = .25 rotate 0, ->- = .75 rotate 0, black!10!orange] plot [smooth, tension=0.5] coordinates { (-1*\bigradius,-1*\gap*3/4)  (0,-1*\gap*3/4) (+1*\bigradius,-1*\gap-1*\gap*3/4)};
        
        \draw [->- = .25 rotate 0, ->- = .75 rotate 0, black!20!cyan] plot [smooth, tension=2] coordinates { (-1*\bigradius,+1*\gap*3/4) (+1*\bigradius,+1*\gap*3/4)};
        
        \node[anchor = east] at (1*\bigradius,1*\bigradius){$\omega$-plane};
    \end{tikzpicture}
	\caption{A generic amplitude $T(s,t)$ when translated to the $\omega$ plane will NOT satisfy $T(\omega,z)^* = T(\omega^*,z)$. Note that the $t$-channel branch cuts of the amplitude unfold (green) along the imaginary $\omega$-axis.}
	\label{fig:omegaplanecut}
\end{figure}
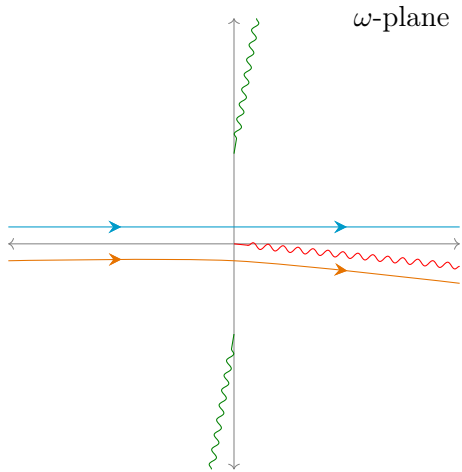

Generally we will be interested in quantities of the form
\begin{equation}
	\int_{0}^{\infty} d \omega\ \omega^{\Delta_{T}-4} \, T(\omega, z)\,, \label{eq:intImT}
\end{equation}
where in general $T(\omega,z)$ could have branch cuts in addition to that of $\omega^{\Delta_T-4}$ (see Figure \ref{fig:omegaplanecut}). Commonly, one relates the imaginary part of the scattering amplitude in the Mandelstam variables $(s,t,u)$
\begin{equation}
\operatorname{Im} T\left(s, t\right)=\frac{1}{2 i}\left(T\left(s, t\right)-{T\left(s, t\right)^*}\right)
\end{equation}
to the corresponding discontinuity across the s-channel cut
\begin{equation}
\operatorname{Disc} T\left(s, t\right)=\lim _{\varepsilon \rightarrow 0^{+}} \frac{1}{2 i}\left({T}\left(s+i \varepsilon, t\right)-{T}\left(s-i \varepsilon, t\right)\right)
\end{equation}
However, notice that this is based on the Schwartz reflection principle which requires the amplitude $T(s,t)$ in the $s$-plane to be real in an interval on the negative real $s$-axis.\footnote{We can illustrate this with the following prototypical example: Take $f(s) = \ln{(e^{-i \pi} s)}$ with $0 \leq \arg{(s)} <2 \pi$. If we evaluate $s$ on the negative real axis, i.e. $s = e^{i \pi} \bar{s}$ for $\bar{s}$ some positive real number, then $f(e^{i \pi} \bar{s}) = \ln{(\bar{s})}$ is real so the discontinuity and imaginary parts should be related. Indeed, $\Im(f(s))\big|_{s = \bar{s}} = -\pi$, while $\mathrm{Disc}(f(s)) \big|_{s = \bar{s}} = \ln{(e^{-i \pi} \bar{s})} - \ln{(e^{-i \pi} e^{2\pi i}\bar{s})} = - 2 \pi i$ for $\bar{s}$ real, so that $\Im(f(s)) = \frac{1}{2i}\mathrm{Disc}(f(s))$ as expected. Notice that had we defined $f(s) = \ln{(e^{-i n \pi} s)}$ for some other odd integer $n$, the function would not have been real at $s = \bar{s}$ real; and indeed, then the imaginary parts and discontinuity would not have matched.} 

In contrast, in $(\omega,z)$ variables the amplitude $T(\omega,z)$ is generically not real along the negative $\omega$-axis, and as a consequence one cannot simply equate the imaginary part with the discontinuity across some branch cut. One instance where this phenomenon arises is the following: take particles $1$ and $2$ to be incoming and consider possible resonances which manifest themselves as poles of $T(s,t)$ in $s = (p_{1} + p_{2})^{2}$ in the second-sheet of $s$. In the $\omega$-plane this translates to the fact that resonances will appear in the lower-half $\omega$-plane with no conjugate pole in the upper-half $\omega$-plane, thus making $T(\omega,z)$ a complex function of $\omega$, i.e. such that $T(\omega,z)^{*} \neq T(\omega^{*},z)$ (see Figure \ref{fig:omegaplanecut}). Indeed, the $O(N)$ model amplitude \eqref{4pt-omegaz} manifestly provides an example where $T(\omega,z)$ is such that it is never real in an interval in the real $\omega$-axis, and the amplitude has a non-trivial imaginary part on the positive $\omega$-axis despite having no discontinuity.

In general, we will assume $T(\omega,z)$ can be written as a sum of two pieces:
\begin{equation}
	 T(\omega,z) = t_{0}(\omega,z) + t_{\log} (\omega,z)\,.
\end{equation}
Here $t_{0}(\omega,z)$ is defined as a rational piece with no branch cut and being possibly complex through the real axis. The remaining $t_{\log}(\omega,z)$ is defined as the piece which contains the logarithmic branch cuts; we assume it to be a real function of $\omega$ up to the discontinuities arising from said branch cuts.\footnote{This is reminiscent of the expansion of $1$-loop amplitudes in terms of $m$-gon $1$-loop scalar integrals in $D$ dimensions
\begin{equation}
    A_{n}^{1-\mathrm{loop}} = \sum_{i}C_{D}^{(i)}I^{(i)}_{D;n} + \sum_{j}C_{D-1}^{(j)}I^{(j)}_{D-1;n} + \cdots + \sum_{k}C_{2}^{(k)}I^{(k)}_{2;n} + \mathrm{Rational \ Terms},
\end{equation}
with the integrals implying the logarithmic pieces, and the rest contained in the rational terms (see e.g. \cite{Elvang:2013cua} and references within).}

To properly define the expression \eqref{eq:intImT} we need to specify the integration contour by taking the different branch cuts into account. Generically, the amplitude $T(s,t)$ has logarithmic branch cuts in the $s$-channel, $t$-channel, or $u$-channel. Since we will be working with massless theories, let us consider the case where the $s$-channel branch point sits at the origin. Notice that the $i\epsilon$-prescription then instructs us to consider a small gap with the $u$-channel cut as we approach the forward limit, i.e. as $t \to 0 \iff z \to \infty$. Since we will be interested in highlighting the effect of resonances let us take the poles to lie away from the positive real $\omega$-axis; although it will not be difficult to include such poles at positive real $\omega$ and massive branch cuts later as done in \cite{Chang:2021wvv}. Altogether, the generic cases we consider have a branch cut lying on (or rather, slightly below) the positive real axis, loosely referred to as the $s$-channel cut.\footnote{See \cite{Hannesdottir:2022bmo} for an extended discussion of the $i\epsilon$-prescription for the S-matrix, including the effects of resonances.}

\begin{figure}
\begin{minipage}{.5\textwidth}
\centering
\resizebox{\columnwidth}{!}{
    \begin{tikzpicture}[]
        \def\bigradius{3}
        \def\gap{3/10}
        
        \draw[help lines,<->] (-1.25*\bigradius, 0) -- (1.25*\bigradius,0);
        \draw[help lines,<->] (0, -0.5*\bigradius) -- (0, 1.25*\bigradius);
        
        \draw[
            line width=1pt, 
            ->- = .13 rotate 0, 
            ->- = .50 rotate 0,
            ->- = .815 rotate 0,
        ]
            let
                \n1 = {0},
                in (\n1:\bigradius) arc (\n1:180-\n1:\bigradius)
                -- cycle;
        \draw[-,decorate,decoration={snake,amplitude=.4mm,segment length=2mm,post length=1mm}] (+10*\gap,-1*\gap) -- (0,0) node[midway, above] {};
        \node[anchor = east] at (1.25*\bigradius,1.25*\bigradius){$\omega$-plane};    
    
        \node[blue] at (-7*\gap,-2*\gap) {$\times$};
        \node[blue] at (-7*\gap,+2*\gap) {$\times$};
        \node[blue] at (-5*\gap,-2.5*\gap) {$\times$};
        \node[blue] at (-5*\gap,+2.5*\gap) {$\times$};
        
        \node[red] at (-3*\gap,-3*\gap) {$\times$};
        \node[red] at (+3*\gap,-3*\gap) {$\times$};
    \end{tikzpicture}
}
\end{minipage}%
\begin{minipage}{.5\textwidth}
\centering
\resizebox{\columnwidth}{!}{
    \begin{tikzpicture}[]
        \def\bigradius{3}
        \def\gap{3/10}
        
        \draw[help lines,<->] (-1.25*\bigradius, 0) -- (1.25*\bigradius,0);
        \draw[help lines,<->] (0, -0.5*\bigradius) -- (0, 1.25*\bigradius);
        
        \draw[
            line width=1pt, 
            ->- = .13 rotate 0, 
            ->- = .50 rotate 0,
            ->- = .815 rotate 0,
        ]
            let
                \n1 = {0},
                in (\n1:\bigradius) arc (\n1:180-\n1:\bigradius)
                -- cycle;
        \draw[-,decorate,decoration={snake,amplitude=.4mm,segment length=2mm,post length=1mm}] (+ 10*\gap,-1*\gap) -- (0,0) node[midway, above] {};
                
        \node[anchor = east] at (1.25*\bigradius,1.25*\bigradius){$\omega$-plane};
        \node[brown] at (-3*\gap,+3*\gap) {$\times$};
        \node[brown] at (+3*\gap,+3*\gap) {$\times$};
        \node[red] at (-3*\gap,-3*\gap) {$\times$};
        \node[red] at (+3*\gap,-3*\gap) {$\times$};
    \end{tikzpicture}
}
\end{minipage}%
\caption{The analytic structure for $s$-channel resonances. Left: Example of the $\omega$-poles of the momentum space amplitude $T$ in $(\omega,z)$ coordinates. The red poles appear without complex conjugate partners, and denote the resonances of the amplitude. Right: $\omega$-pole structure for the imaginary part $(T(\omega,z) - T^{*}(\omega,zd))/2i$. We have used the $O(N)$ model to exemplify the analytic structure.}
\label{fig:analyticSChannel}
\end{figure}
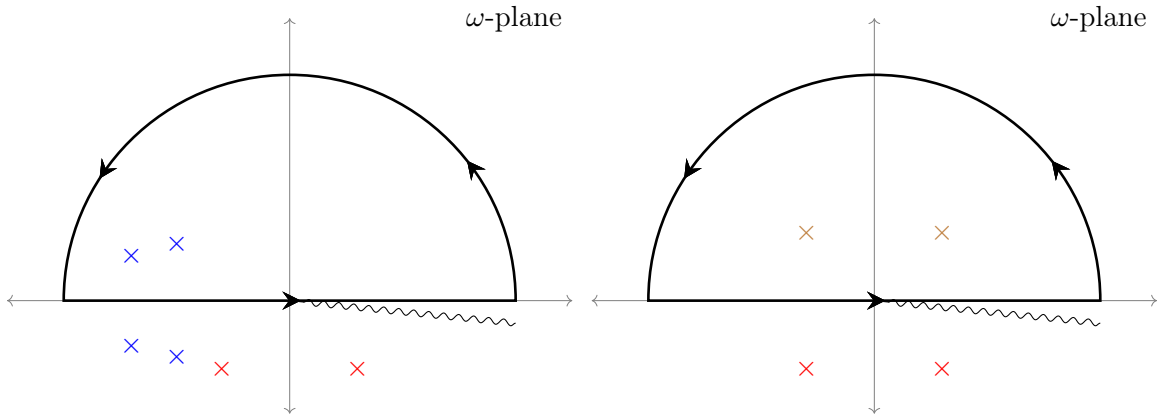

In order to see the effect of resonances on the imaginary part of the celestial amplitude let us focus on the case where $t_{0}(\omega,z)$ fails to be a real function of $\omega$ because of these resonances. For instance, taking particles 1 and 2 to be incoming, there are poles of $t_{0}(\omega,z)$ in the lower-half $\omega$-plane which appear with no complex conjugate. These are the $s$-channel resonances of $T(\omega,z)$, which we label as $\omega_{\mathrm{res}}$. All other poles come then in conjugate pairs with conjugate residues. An illustration of this analytic structure is depicted in Figure \ref{fig:analyticSChannel}. 

To compute the Mellin-transform of $t_{0}(\omega,z)$, we can proceed by extending the integral to the whole real line, and then deforming the contour to pick the residues in the upper $\omega$-plane. The integral has a branch cut beginning at $\omega=0$, for which we choose the branch cut as depicted in Figure \ref{fig:analyticSChannel}. The extension of the integral to the whole real line is then, explicitly:
\begin{align} 
    \int_{0}^{\infty} d\omega \, \omega^{\Delta_{T}-4} \, t_{0}(\omega,z) & = \frac{1}{(1 + e^{i \pi \Delta_{T}})} \int_{-\infty}^{\infty} d\omega \, \omega^{\Delta_{T} - 4} \, \frac{t_{0}(\omega,z) + t_{0}(-\omega,z)}{2} \nonumber \\[0.3cm] &  + \frac{1}{(1 - e^{i \pi \Delta_{T}})} \int_{-\infty}^{\infty} d\omega \, \omega^{\Delta_{T} - 4} \, \frac{t_{0}(\omega,z) - t_{0}(-\omega,z)}{2} \nonumber \\[0.3cm]
    & = \frac{i e^{-i \pi \Delta_{T}}}{4 \sin{(\frac{\pi \Delta_{T}}{2})} \cos{(\frac{\pi \Delta_{T}}{2})}  } \int_{-\infty}^{\infty} d\omega \, \omega^{\Delta_{T} - 4} \, t_{0}(\omega,z) \nonumber \\[0.3cm] & - \frac{i }{4 \sin{(\frac{\pi \Delta_{T}}{2})} \cos{(\frac{\pi \Delta_{T}}{2})}  } \int_{-\infty}^{\infty} d\omega \, \omega^{\Delta_{T} - 4} \, t_{0}(-\omega,z)\,. \label{extensionformula}
\end{align}

Specializing to the imaginary part $\mathrm{Im}(t_{0})(\omega,z) = (t_{0}(\omega,z) - t_{0}^{*}(\omega,z))/2i$ we find poles in both the lower-half and upper-half $\omega$-plane, at $\omega_{\mathrm{res}}$ and $\omega^{*}_{\mathrm{res}}$, arising from the $t_{0}(\omega,z)$ and $t_{0}^{*}(\omega,z)$ terms respectively. Any other poles that could have appeared conjugated cancel when taking the imaginary part (see Figure \ref{fig:analyticSChannel}). In other words, the poles corresponding to resonances from the second-sheet appear in the lower-half $\omega$-plane and without complex conjugate partners. When looking at the imaginary part of the amplitude these contributions do not disappear, the way poles not in the scattering channel would, and so provide residues for the contour integral in \eqref{extensionformula}.\footnote{It is interesting to note that if we had poles on the first-sheet of the S-matrix (which generally indicates issues with causality of the scattering amplitude), they would turn into non-complex conjugate partners in the \textit{upper half} $\omega$-plane, and so would contribute to the contour integrals with the wrong-sign residues.}

Since any poles appearing with their conjugate in $t_{0}(\omega,z)$ have been cancelled in $\mathrm{Im}(t_{0})(\omega,z)$, we can restrict to only those poles that are associated with resonances. We do this in the following. Applying \eqref{extensionformula} to $\mathrm{Im}(t_{0})(\omega,z)$ instead of $t_{0}(\omega,z)$, we see that in the first term of the last expression only $-t_{0}^{*}(\omega,z)$ contributes since $t_{0}(\omega,z)$ has no poles in the upper $\omega$-plane, and similarly in the second term $\mathrm{Im}(t_{0})(-\omega,z)$ where only $t_{0}(-\omega,z)$ contributes (the poles of $t_{0}^{*}(-\omega,z)$ now appearing in the lower $\omega$-plane). Altogether, we obtain the contribution of $t_{0}(\omega,z)$ to the imaginary part of the celestial amplitude as
\begin{equation}
    \frac{- i \pi}{4 \sin{(\frac{\pi \Delta_{T}}{2})} \cos{(\frac{\pi \Delta_{T}}{2})}} \sum_{\omega_{\mathrm{res}}} \bigg[ e^{-i \pi \Delta_{T}} \, \underset{\omega = \omega^{*}_{\mathrm{res}}}{\mathrm{\mathbf{Res}}}\Big(\omega^{\Delta_{T} - 4} t_{0}^{*}(\omega,z) \Big) + \underset{\omega = - \omega_{\mathrm{res}}}{\mathrm{\mathbf{Res}}}\Big(\omega^{\Delta_{T} - 4} t_{0}(-\omega,z) \Big) \bigg]\,, \label{t0celestialcontribution}
\end{equation}
which is solely determined by the residues at the $s$-channel resonances. In particular, when $t_{\log}(\omega,z) = 0$, the imaginary part of the celestial amplitude is fully determined by the $s$-channel resonances, and is equal to the above expression.

Interestingly, the previous expression allows us to consider the question of how to identify resonances directly from the celestial perspective, with no reference to the more standard momentum space. Namely, expand the imaginary part of the celestial amplitude in the form \eqref{t0celestialcontribution} and read-off the $\Delta_{T}$ dependence. This should be given by contributions that schematically depend as $(\omega_{\mathrm{res}}^{*})^{\Delta_{T}-4}$ or $(-\omega_{\mathrm{res}})^{\Delta_{T}-4}$ with $\omega_{\mathrm{res}}^{*}$ or $-\omega_{\mathrm{res}}$ having a positive imaginary part, recalling that $\omega_{\mathrm{res}}$ corresponds to a resonance that in the $\omega$-plane lied in the lower-half plane. The resonances can then be identified as the conjugates of those contributions providing the non-trivial $\Delta_{T}$ dependence in \eqref{t0celestialcontribution}. At the end of this section we will illustrate the previous point using the $O(N)$ model amplitude as an example.

If we also consider the logarithmic pieces, we can write the imaginary part of the celestial amplitude as:
\begin{align}
    \Im(\mathcal{A}\big( \Delta_{T},z) \big) =  
        &\frac{- i \pi}{4 \sin{(\frac{\pi \Delta_{T}}{2})} \cos{(\frac{\pi \Delta_{T}}{2})}} \nonumber \\
        &\times\sum_{\omega_{\mathrm{res}}} \bigg[ e^{-i \pi \Delta_{T}} \, \underset{\omega = \omega^{*}_{\mathrm{res}}}{\mathrm{\mathbf{Res}}}\Big(\omega^{\Delta_{T} - 4} t_{0}^{*}(\omega,z) \Big) + \underset{\omega = - \omega_{\mathrm{res}}}{\mathrm{\mathbf{Res}}}\Big(\omega^{\Delta_{T} - 4} t_{0}(-\omega,z) \Big) \bigg]\nonumber\\ 
        &+  \int_0^\infty d\omega\,\omega^{\Delta_T-4}  \big[ \mathrm{Disc}(t_{\log})\big](\omega,z) \label{eq1}\,.
\end{align}
If instead of a branch cut at the origin we have a discrete series of simple poles at positive masses $\omega = m_{i}$ and a massive branch cut starting at $\omega = M$ described by a real function of $\omega$ (up to the discontinuities implied by such branch cut as in \cite{Chang:2021wvv}), then we just replace the last integral in \eqref{eq1} by a discontinuity integral starting at $M$ instead and a series of residues at $\omega = m_{i}$. That is, the result of \cite{Chang:2021wvv} is recovered, but with contributions coming from resonances added (indeed, as we will see next, the imaginary part of the $O(N)$ model is fully explained by the latter, with no contributions arising either from discontinuities or poles at positive masses $\omega = m_{i}$). 

We see that the imaginary part of a celestial amplitude captures both ``the data at the positive real $\omega$-axis'' such as discontinuities or residues of poles at positive masses (had we considered massive theories), as well as the data from resonances at the lower-half $\omega$-plane through non-trivial contributions.

Let us illustrate the previous discussion for the case of the $O(N)$ model. Since we are considering $12 \rightarrow 34$ scattering, the contributions proportional to $\delta^{ac}\delta^{bd}$ and $\delta^{ad}\delta^{bc}$ have complex conjugate poles (depicted blue in Figure \ref{fig:analyticSChannel}) and by the previous discussion we know they will not play a role in what follows.\footnote{Similarly, if we were considering physical $13 \rightarrow 24$ ($14 \rightarrow 23$) scattering, then $t > 0$ ($u>0$) and the $\delta^{ac}\delta^{bd}$ ($\delta^{ad}\delta^{bc}$) term would have the resonance poles in the lower-half $\omega$-plane. The remaining terms would have conjugate poles.} The $O(N)$ model is also a case where $t_{\log}(\omega,z) = 0$ and where no massive poles appear in the real positive axis for finite values of the coupling $\lambda$, so, we can cleanly concentrate on the non-trivial aspects of resonances only. 

The previous paragraph implies we only need to consider the contribution proportional to $\delta^{ab}\delta^{cd}$:
\begin{equation}\label{ab-cdpart}
    \frac{\omega^{2}}{(\omega^{2} + i \frac{\lambda}{16}\omega + 2 \mu^{2})}\delta^{ab}\delta^{cd} = \frac{\omega^{2}}{{(\omega - \omega_{+}})(\omega - \omega_{-})}\delta^{ab}\delta^{cd}\,,
\end{equation}
where the $s$-channel resonances of $T^{ab,cd}_{4}(\omega,z)$ --which lie in the lower-half $\omega$-plane-- are the $\omega^{\pm}_{ab,cd}$ poles in \eqref{omegaabcd}, which we have written here merely as $\omega_{\pm} \coloneqq \omega^{\pm}_{ab,cd}$ for simplicity. These appear depicted in red in Figure \ref{fig:analyticSChannel}. 

We could now directly apply \eqref{t0celestialcontribution}, but let us notice that the $O(N)$ model is a case where \eqref{ab-cdpart} satisfies $T^{*}(\omega,z) = T(-\omega,z)$. In general, when this condition is satisfied for some $T(\omega,z)$ we have that
\begin{equation}
    \sum_{\omega_{\mathrm{res}}} \underset{\ \omega = \omega^{*}_{\mathrm{res}}}{\mathrm{\mathbf{Res}}}\Big(\omega^{\Delta_{T} - 4} T^{*}(\omega,z) \Big) = \sum_{\omega_{\mathrm{res}}} \underset{\ \omega = -\omega_{\mathrm{res}}}{\mathrm{\mathbf{Res}}} \Big(\omega^{\Delta_{T} - 4} T(-\omega,z) \Big)\,,
\end{equation}
in which case Equation \eqref{t0celestialcontribution} simplifies to
\begin{equation}
\boxed{
    \mathrm{Im}\big(\mathcal{A}(\Delta_{T},z)\big) = - \frac{\pi}{(1-e^{\pi i \Delta_{T}})} \sum_{ \omega_{\mathrm{res}}} \underset{\ \omega = \omega^{*}_{\mathrm{res}}}{\mathrm{\mathbf{Res}}} \Big( \omega^{\Delta_{T}-4} T^{*}(\omega,z) \Big)\,.} \label{simplifiedoptical}
\end{equation}
In particular, this expression can be used to more directly evaluate $\mathrm{Im}(\mathcal{A})$ in the case of the $O(N)$ model.
The relevant residues of $T^{*}(\omega,z)$ for the $O(N)$ model are located at
\begin{equation}
    \omega_{\pm}^{*} = \sqrt{-2\mu^{2}} e^{i \pi/2} e^{\pm i \theta} \, ,
\end{equation}
and are straightforwardly evaluated to verify that:
\begin{equation}
    \underset{\ \omega = \omega_{\pm}^{*}}{\mathrm{\mathbf{Res}}} \big(\omega^{\Delta_{T} - 4} T^{*}(\omega) \big) = -\frac{(e^{-i \pi} \omega_{\mp})^{\Delta_{T} - 4}}{2\omega_{\mp}}\underset{\ \omega^2 = \omega_{\mp}^2}{\mathrm{\mathbf{Res}}} \big( T(\omega) \big) = \frac{(e^{-i\pi} \omega_{\mp})^{\Delta_{T} - 4}}{2 \omega_{\mp}} T^{ab}_{3}(\omega_{\mp}) T^{cd}_{3}(\omega_{\mp})\,.
\end{equation}
In this expression the first equality stresses the point that the imaginary part of the celestial amplitudes arises from resonances going on-shell. This is also recognized in the last equality, which has been written as a product of the three-point function \eqref{eq:3point} satisfying \eqref{fourpointfromthreepointsquare}.\footnote{Up to a Wick-rotation taking us from Euclidean to Lorentzian signature.} Obviously, this leads to the imaginary part of the celestial amplitude:
\begin{align}
   \mathrm{Im}\big(\mathcal{A}(\Delta_{T},z)\big) &= - \frac{\pi e^{-i\pi(\Delta_{T}-4) }}{2(1-e^{\pi i \Delta_{T}})} \bigg[ \omega_{+}^{\Delta_{T} - 5} T^{ab}_{3}(\omega_{+}) T^{cd}_{3}(\omega_{+}) + \omega_{-}^{\Delta_{T} - 5} T^{ab}_{3}(\omega_{-}) T^{cd}_{3}(\omega_{-}) \bigg]\label{eq:AmplitudeFactorization}
    \\[0.25cm]  &= -\Big(\frac{\pi \lambda}{2 N}\Big)(-2\mu^{2})^{(\Delta_{T}-3)/2} \frac{\sin{[(\Delta_{T}-2)\theta]}}{\sin{\theta}\sin{(\frac{\pi \Delta_{T}}{2})}} \, \delta^{ab} \delta^{cd}, \label{impart}
\end{align}
fully determined by the resonances of the theory going on-shell. Note: To obtain the correct pre-factors for the imaginary part of the celestial amplitude it was crucial that both resonances appeared in the sum. This fact will play a role for us again when we study the four-point functions decomposition into conformal blocks and factorization into three-point coefficients in Section \ref{sec:factorization}.

Let us note that when considering the limit $\lambda \rightarrow 0$, formula \eqref{simplifiedoptical} leads to two contributions at $\omega = \pm \sqrt{-2\mu^{2}} + \mathcal{O}(\lambda)$. These mix together to cancel the monodromy factor in \eqref{simplifiedoptical}, obtaining:
\begin{equation}
    \lim_{\lambda \rightarrow 0} \mathrm{Im}\Big(\mathcal{A}(\Delta_{T},z)\Big) = \pi  \underset{\ \omega = -2\mu^{2}}{\mathrm{\mathbf{Res}}} \bigg( \omega^{\Delta_{T}-4} T(\omega,z) \Big|_{\lambda=0} \bigg),
\end{equation}
in accordance with the non-trivial imaginary part at tree-level as considered in \cite{Chang:2021wvv}.

In the converse direction, let us illustrate how to locate the resonances of the $O(N)$ model directly from the celestial amplitude following the discussion in the paragraph right below Equation \eqref{t0celestialcontribution}. That is, suppose we are handed the celestial amplitude \eqref{Mtransform} and we are asked to check for the presence of resonances. According to the previous discussion, we can take the imaginary part given by Equation \eqref{impart}, and expand it in the form \eqref{t0celestialcontribution}. A quick bit of algebra shows that:
\begin{align}
   &\mathrm{Im}\big(\mathcal{A}(\Delta_{T},z)\big) = \frac{-i\pi \lambda \delta^{ab} \delta^{cd}}{8N \sin{(\frac{\pi \Delta_{T}}{2})} \cos{(\frac{\pi \Delta_{T}}{2})}}  \frac{(-2\mu^{2})^{1/2}}{\sin{\theta}} \bigg\{ e^{-i \pi \Delta_{T}}\Big[(\sqrt{-2\mu^{2}} e^{i(\frac{\pi}{2} - \theta)})^{\Delta_{T}-4} e^{-2i\theta} \nonumber
    \\[0.25cm]  &- (\sqrt{-2\mu^{2}} e^{i(\frac{\pi}{2} + \theta)})^{\Delta_{T}-4} e^{2i\theta}\Big] + \Big[(\sqrt{-2\mu^{2}} e^{i(\frac{\pi}{2} - \theta)})^{\Delta_{T}-4} e^{-2i\theta}- (\sqrt{-2\mu^{2}} e^{i(\frac{\pi}{2} + \theta)})^{\Delta_{T}-4} e^{2i\theta}\Big] \bigg\}.
\end{align}
Clearly, the different summands carry factors of the form $\omega_{*}^{\Delta_{T}-4}$ with $\omega_{*}$ having a positive imaginary part (e.g. $\omega_{*}=(\sqrt{-2\mu^{2}} e^{i(\frac{\pi}{2} - \theta)}$). Following our previous discussion the position of the resonances can then be read off as the conjugates, and as obviously expected they correspond to the resonances \eqref{omegaabcd}, although here we have obtained them from the celestial amplitude instead of the momentum space amplitude.

\section{Conformal Decompositions}\label{sec:confDec}
Studying the conformal partial wave decompositions of 2d celestial amplitudes has been of great interest in elucidating the optical theorem, soft limits, light ray operators, and more (see e.g. \cite{lamShao, celestialAmplitudes, Fan:2021isc, Atanasov:2021cje, Chang:2021wvv}). In this section, we start by looking at the conformal block decompositions of our 1d celestial amplitudes (assuming all external scaling dimensions are the same for simplicity). We find that there are two infinite towers of operators: ``single-trace'' at exchange dimension $\Delta_{\calO} = 2n+1$, coming (roughly) from the kinematic pre-factor of the celestial amplitude; and ``double-trace'' at exchange dimension $\Delta_{\calO} = \alpha+n+1$, where $n\in\bbN^0$. 

Focusing on the imaginary part of the amplitude, we verify that we can reproduce the coefficients of the conformal block decomposition by the related conformal partial wave decomposition. The $J=1$ series of antisymmetric conformal partial waves appear because we consider four-point functions with colour. We see the cancellations that occur between the discrete series conformal partial waves and spurious poles obtained via the continuous series.

An important question is to what extent the conformal block coefficients in the expansions can be viewed as ``three-point function coefficients squared.'' We argue that for the single-trace operators, the fact that there are two poles on the second sheet of the $p^2$-plane in the $O(N)$ model is directly relevant for this to hold. We briefly comment on the extent to which the double-trace operators can be factorized.

Background on correlation functions in 1d CFT, conformal blocks, and conformal partial waves appear in Appendix \ref{app:CFTBackgroundConventions}.

\subsection{Color Structure of Celestial Amplitudes}\label{sec:celestialAmps}
The celestial amplitudes in Equations \eqref{eq:CelestialONfu} - \eqref{eq:CelestialONfs} sew together piecewise into a reasonable 1d CFT correlator of bosons (see Appendix \ref{sec:1dCFTGeneral} for properties of 1d CFT correlators), given by
\begin{equation}
    f^{abcd}(z) \coloneqq
        \begin{cases}
            f_u^{abcd}(z) & \text{if $z\in(-\infty,0)$}\\
            f_t^{abcd}(z) & \text{if $z\in(0,1)$}\\
            f_s^{abcd}(z) & \text{if $z\in(1,\infty)$}
        \end{cases}\,. \label{eq:fstu}
\end{equation}

We may further decompose $f^{abcd}(z)$ into its $O(N)$ representations. We define $O(N)$ color tensors
\begin{equation}
    T_S^{abcd}=\delta^{ab}\delta^{cd}, \quad
    T_A^{abcd}=\delta^{ac}\delta^{bd}-\delta^{ad}\delta^{bc}, \quad
    T_T^{abcd}=\delta^{ac}\delta^{bd}+\delta^{ad}\delta^{bc}-\frac{2}{N}\delta^{ab}\delta^{cd},\label{eq:colorTensors}
\end{equation}
labelling symmetric, antisymmetric, and traceless components respectively, and corresponding components $f_{I}(z)$ for $f^{abcd}(z)$, i.e.
\begin{equation}
    f^{abcd}(z) = f_S(z) T_S^{abcd} + f_A(z) T_A^{abcd} + f_T(z) T_T^{abcd}\,.
\end{equation}

We might call the choice of definitions in Equation \eqref{eq:colorTensors} the ``$s$-channel color tensors'' since, for example, $T_S^{abcd}$ really only looks like the singlet when scattering $12\to34$. In decomposing $t$-channel amplitudes, we instead use the basis given by $T_S^{acbd}$, $T_A^{acbd}$, $T_T^{acbd}$, and similarly for $u$. i.e.
\begin{alignat}{4}
    f_{u}^{abcd}(z) 
        &= f_{u,S}(z) T_S^{adcb} &&+ f_{u,A}(z) T_A^{adcb} &&+ f_{u,T}(z) T_T^{adcb}\,,\\
    f_{t}^{abcd}(z) 
        &= f_{t,S}(z) T_S^{acbd} &&+ f_{t,A}(z) T_A^{acbd} &&+ f_{t,T}(z) T_T^{acbd}\,,\\
    f_{s}^{abcd}(z) 
        &= f_{s,S}(z) T_S^{abcd} &&+ f_{s,A}(z) T_A^{abcd} &&+ f_{s,T}(z) T_T^{abcd}\,.
\end{alignat}

If we write out the celestial amplitudes in components we see that the ``imaginary pieces'' that show up in the optical theorem, distinguished by the $e^{i\pi\alpha}$ factors, find themselves entirely in the color-singlet channel:
\begin{align}
    f_S(z) &= \frac{\calN \calN_{\alpha}}{N} \frac{\abs{z}}{\sqrt{\abs{1-z}}}\times
        \begin{cases}
            1 + (-z)^{\alpha} + Ne^{i\pi\alpha}\left(\frac{-z}{1-z}\right)^{\alpha} 
                & \text{if $z\in(-\infty,0)$}\\
            1 + Ne^{i\pi\alpha}z^{\alpha} + \left(\frac{z}{1-z}\right)^{\alpha}
                & \text{if $z\in(0,1)$}\\
            Ne^{i\pi\alpha} + z^{\alpha} + \left(\frac{z}{z-1}\right)^{\alpha} 
                & \text{if $z\in(1,\infty)$}
        \end{cases}\,,  \\[0.5cm]
    f_A(z) &= \frac{\calN \calN_{\alpha}}{2} \frac{\abs{z}}{\sqrt{\abs{1-z}}}\times
        \begin{cases}
            -1 + (-z)^{\alpha} 
                & \text{if $z\in(-\infty,0)$}\\
            1 - \left(\frac{z}{1-z}\right)^{\alpha}
                & \text{if $z\in(0,1)$}\\
            z^{\alpha} - \left(\frac{z}{z-1}\right)^{\alpha} 
                & \text{if $z\in(1,\infty)$}
        \end{cases}\,, \\[0.5cm]
    f_T(z) &= \frac{\calN \calN_{\alpha}}{2} \frac{\abs{z}}{\sqrt{\abs{1-z}}}\times
        \begin{cases}
            1 + (-z)^{\alpha} 
                & \text{if $z\in(-\infty,0)$}\\
            1 + \left(\frac{z}{1-z}\right)^{\alpha}
                & \text{if $z\in(0,1)$}\\
            z^{\alpha} + \left(\frac{z}{z-1}\right)^{\alpha} 
                & \text{if $z\in(1,\infty)$}
        \end{cases}\,.
\end{align}
We've assumed all external scaling dimensions $\Delta_i$ are the same above. For comparison, the four-point function of the $O(N)$ generalized free boson with scaling dimension $\Delta_{\mathrm{GFB}}$ is obtained by removing the kinematic prefactors, removing the exponential factors $e^{i\pi\alpha}$, and replacing $\alpha \mapsto 2\Delta_{\mathrm{GFB}}$. It can be decomposed in the same way.

\subsection{Conformal Block Decomposition}\label{sec:confBlock}
We start by considering the conformal block decomposition of $f_t^{abcd}(z)$ where $z\in(0,1)$. The components in the singlet, anti-symmetric, and symmetric traceless reps respectively are
\begin{alignat}{3}
    f_{t,S}(z) 
        &= \frac{\calN \calN_{\alpha}}{N} \frac{z}{\sqrt{1-z}} 
        &&\left(1 + N e^{i\pi\alpha} z^\alpha + \left(\frac{z}{1-z}\right)^{\alpha}\right)\,,\\
    f_{t,A}(z) 
        &= \frac{\calN \calN_{\alpha}}{2} \frac{z}{\sqrt{1-z}} 
        &&\left(1 - \left(\frac{z}{1-z}\right)^{\alpha}\right)\,,\\
    f_{t,T}(z) 
        &= \frac{\calN \calN_{\alpha}}{2} \frac{z}{\sqrt{1-z}} 
        &&\left(1 + \left(\frac{z}{1-z}\right)^{\alpha}\right)\,.
\end{alignat}

To proceed with a conformal block decomposition, we just need the conformal block decomposition of some common $z$-dependent pieces. In Section 2 of \cite{Hogervorst:2017sfd} the authors give the conformal block decompositions we need as
\begin{equation}
    z^p (1-z)^{-q} = \sum_{n=0}^{\infty} c_{p,q}(n) k_{p+n}(z)\,, \qquad 0<z<1\,,
\end{equation}
where $k_{\Delta}$ is the $SL(2,\mathbb{R})$ conformal block with all external $\Delta_i$ the same,\footnote{In a CCFT, where scaling dimensions are continuous, taking all external dimensions to be the same could be a large loss of information. However, we make this specialization so that we can obtain closed form expressions in terms of hypergeometrics and not just series expansions. Using external scaling dimensions with $\Delta_{12} = \Delta_{34} = 0$ isn't much harder (see footnote \ref{foot:unwrittenDeltas}).} i.e.
\begin{equation}
    k_{\Delta}(z) = z^\Delta {}_2F_1(\Delta,\Delta;2\Delta;z)
\end{equation}
(see also Appendix \ref{app:confBlock}). The function $c_{p,q}(n)$ is defined as
\begin{equation}
    c_{p,q}(n) \coloneqq \frac{(p)^2_n}{n! (2p-1+n)_n} \pFq{3}{2}{p-q,2p-1+n,-n}{p,p}{1}\,. \label{eq:defcpq}
\end{equation}
We note that the three quantities relevant to us satisfy\footnote{To see the last relation, one can manipulate gamma functions, or use the following identity for $n$ a positive integer (see Sections 3.5 and 3.9 of \cite{bailey1935generalized})
\begin{equation}
    \pFqreg{3}{2}{a,b,-n}{e,f}{1} = (-1)^n \frac{\Gamma(1-f+b)}{\Gamma(f+n)} \pFqreg{3}{2}{e-a,b,-n}{e,1-f+b-n}{1}\,.
\end{equation}}
\begin{align}
    c_{1,\frac{1}{2}}(2k+1) 
        &= 0 \,,\\
    (-1)^nc_{\alpha+1,\frac{1}{2}}(n) &= c_{\alpha+1,\alpha+\frac{1}{2}}(n)\,.\label{eq:coefficientIdentity}
\end{align}

Putting all the pieces together, we can write $f_{t,I}$ as a sum over two towers of exchange operators
\begin{equation}\label{eq:conformalBlockTowers}
    \frac{1}{\calN \calN_{\alpha}} f_{t,I}(z) 
        = \sum_{n=0}^\infty B_I(n) k_{2n+1}(z) 
        + \sum_{n=0}^\infty C_I(n) k_{n+\alpha+1}(z)
\end{equation}
with
\begin{equation}
    N B_S(n) =
    2 B_A(n) = 
    2 B_T(n) = 
    c_{1,\frac{1}{2}}(2n) = 
    \frac{16^n \,\Gamma (\frac{1}{2}+n)^2}{\Gamma(4n+1)\,\Gamma(\frac{1}{2}-n)^2}\,,
\end{equation}
and
\begin{equation}
    C_I(n) = 
    c_{\alpha+1,\alpha+\frac{1}{2}}(n)
    \times
    \begin{cases}
        \frac{1}{N}(1+(-1)^nNe^{i\pi\alpha}) & \text{if $I=S$}\\    
        -\frac{1}{2} & \text{if $I=A$}\\    
        \phantom{-}\frac{1}{2} & \text{if $I=T$}\\    
    \end{cases}\,.
\end{equation}
All-in-all, we see that there are ``single-trace operators'' exchanged in the $\phi\times\phi$ OPE with odd-integer dimension,\footnote{When all $\Delta_i$ are not the same, the exchanges at even integer dimension don't vanish. In other words, there is no special mechanism, besides our specificity of external scaling dimension, that makes the $\Delta_{\calO}=2n$ contributions disappear. \label{foot:unwrittenDeltas}} i.e. no dependence on the scaling dimensions of the external particles; and also operators that look more like familiar ``double-trace'' operators
\begin{equation}
    \Delta_{\calO} = 2n+1\,,\qquad 
    \Delta_{\calO} = \alpha+n+1 = 2\Delta_{\phi}+n-\frac{1}{2}\,,\qquad
    n\in\bbN^0\,.
\end{equation}


Recall that for $z\in(0,1)$ there are actually two convergent OPE channels, the $z\to0$ channel and the $z\to1$ channel. Our decomposition in Equation (\ref{eq:conformalBlockTowers}) is for the $z\to 0$ channel, but it's not hard to see that this decomposition holds for the crossed channel coming from the $z\to 1$ OPE. In other words, we have that
\begin{equation}
    \frac{1}{\calN \calN_{\alpha}} \left(\frac{1-z}{z}\right)^{\alpha+3/2} f_{t,I}(z) 
        = \sum_{n=0}^\infty B_I(n) k_{2n+1}(1-z) 
        + \sum_{n=0}^\infty C_I(n) k_{\alpha+n+1}(1-z)
\end{equation}
for $I=S,T$, and the same with an overall minus sign for $I=A$ from swapping $1$ and $3$ in the OPE. Of course, this is trivially true because $f_{S,T}(z)=f_{S,T}(1-z)$ and $f_A(z)=-f_A(1-z)$, but it is nice to check the decompositions.


\subsubsection{The Imaginary Pieces}
We might also consider just the ``imaginary piece'' of the celestial amplitude, $\Im(f^{abcd}(z))$, which is also a well-defined correlator of bosons (see Equations \eqref{eq:commRln1}-\eqref{eq:commRln3}). It is given by
\begin{equation}
    \Im(f^{abcd}(z)) = \sin({\pi \alpha})\,\calN \calN_{\alpha} \frac{\abs{z}}{\sqrt{\abs{1-z}}}\times
        \begin{cases}
            \delta^{ad}\delta^{bc} \left(\frac{-z}{1-z}\right)^{\alpha}
                & \text{if $z\in(-\infty,0)$}\\
            \delta^{ac}\delta^{bd} z^{\alpha}  
                & \text{if $z\in(0,1)$}\\
            \delta^{ab}\delta^{cd}  
                & \text{if $z\in(1,\infty)$}
        \end{cases} \label{eq:imaginaryAmplitude}\,.
\end{equation}

Re-purposing the computations from above, we have the CFT crossing-symmetric results
\begin{align}
    \frac{\csc({\pi \alpha})}{\calN \calN_{\alpha}}\Im(f_u^{abcd}(z)) 
        &= \delta^{ad}\delta^{bc} \sum_{n=0}^{\infty} c_{\alpha+1,\frac{1}{2}}(n) k_{\alpha+n+1}\left(\frac{z}{z-1}\right)\,,\\
    \frac{\csc({\pi \alpha})}{\calN \calN_{\alpha}}\Im(f_t^{abcd}(z)) 
        &= \delta^{ac}\delta^{bd} \sum_{n=0}^{\infty} c_{\alpha+1,\frac{1}{2}}(n) k_{\alpha+n+1}(z)\,,\\
    z^{-(\alpha+\frac{3}{2})}\frac{\csc({\pi \alpha})}{\calN \calN_{\alpha}}\Im(f_s^{abcd}(z)) 
        &= \delta^{ab}\delta^{cd} \sum_{n=0}^{\infty} c_{\alpha+1,\frac{1}{2}}(n) k_{\alpha+n+1}\left(\frac{1}{z}\right)\,.
\end{align}
We see that the imaginary pieces, which play a role in the optical theorem, contribute to the tower of double-exchange operators that we saw earlier.

\subsection{Conformal Partial Waves: Euclidean Inversion of \texorpdfstring{$\Im(f)$}{Im(f)}}\label{sec:CPWs}
Here we explicitly apply the Euclidean inversion formula to the imaginary piece of the celestial amplitude $\Im(f^{abcd}(z))$ in an attempt to reproduce the conformal block decomposition answer. We also see the nice cancellations that occur with the discrete series. Our conventions, definitions, and identities for 1d conformal partial waves are in Appendix \ref{sec:CPWFormulas}.

In the case of the Euclidean inversion formula, we just naively apply the inner product to Equation (\ref{eq:imaginaryAmplitude}). We will temporarily put aside the contributions from the discrete series $\tilde{I}_{m}^{abcd}$ and focus on the integrals to do in computing
\begin{equation}
    I_{\Delta,J}^{abcd} \coloneqq (\Psi_{\Delta,J},\Im f^{abcd} ) = \int_{-\infty}^{\infty} \frac{dz}{z^2}  \Im(f^{abcd})(z) \Psi_{\Delta,J}(z)
\end{equation}
along the principal continuous series conformal partial waves, $\Psi_{\Delta,J}(z)$.

It is easiest to start with the integral over the $z\in(0,1)$ region. For this, we compute
\begin{align}
    I^{(0)}_{\Delta,J} 
        &\coloneqq \int_{0}^1 \frac{dz}{z^2} \frac{z^{\alpha+1}}{\sqrt{1-z}}\Psi^{(0)}_{\Delta, J}(z)\\
        &= \sqrt{\pi} \left((-1)^J+\sec (\pi \Delta)\right) \Gamma(\Delta)^2 \Gamma(\alpha+\Delta)\, \pFqreg{3}{2}{\Delta,\Delta,\alpha+\Delta}{2\Delta,\alpha+\Delta+\frac{1}{2}}{1}\\
        &+ (\Delta \leftrightarrow 1-\Delta)\nonumber\,.
\end{align}
The integral over the $z\in(-\infty,0)$ region can be completed by using the identity in Equation \eqref{eq:PsiMinusFromZero}
\begin{equation}
    I^{(-)}_{\Delta,J} 
        \coloneqq \int_{-\infty}^0 \frac{dz}{z^2} \frac{(-z)^{\alpha+1}}{(1-z)^{\alpha+\frac{1}{2}}}\Psi^{(-)}_{\Delta, J}(z)
        = \int_{0}^1 \frac{dw}{w^2} \frac{w^{\alpha+1}}{\sqrt{1-w}}(-1)^J \Psi^{(0)}_{\Delta, J}(w)\,,
\end{equation}
so that $I^{(-)}_{\Delta,J} = (-1)^J I^{(0)}_{\Delta,J}$.

For the integral over $z\in(1,\infty)$, one must use the integral form of the partial wave and break up the absolute values over different regions. Luckily, we can instead reuse the result from Equation (4.20) of \cite{lamShao}, which tells us that
\begin{equation}
     I_{\Delta,0}^{(+)}
        \coloneqq \int_{1}^{\infty} \frac{dz}{z^2} \frac{z}{\sqrt{z-1}} \Psi^{(+)}_{\Delta,0}(z)\\
        = \frac{4\pi^3}{\sin(\pi \Delta) \Gamma\left(\frac{2-\Delta}{2}\right)^2\Gamma\left(\frac{\Delta+1}{2}\right)^2}\,.
\end{equation}
Because $\Im(f_s^{abcd}(z))$ is symmetric under $z\mapsto \frac{z}{z-1}$, the $J=1$ integral vanishes and $I^{(+)}_{\Delta,1} = 0$.

We can now decompose the four-point function over the conformal partial waves using the Euclidean inner product (see Equation \eqref{eq:1dpartialWaveDecomposition} and surrounding discussion), and shadow symmetry of the various components, to write
\begin{equation}
    \Im(f^{abcd})(z) 
        = \sum_{J=0,1}\int_{\frac{1}{2}-i\infty}^{\frac{1}{2}+i\infty} \frac{d\Delta}{2\pi i} \frac{I^{abcd}_{\Delta,J}}{2n_{\Delta}} \Psi_{\Delta,J}(z) 
        + \sum_{m=0}^{\infty} \frac{2m-1}{4\pi^2} \tilde{I}^{abcd}_m \Psi_{m,J}(z)\,,
\end{equation}
where the combined coefficient is
\begin{equation}
    I^{abcd}_{\Delta,J} 
        = \sin(\pi\alpha)\, {\calN}\, {\calN}_{\alpha} 
        \left(
            (-1)^J \delta^{ad}\delta^{bc} I_{\Delta,J}^{(0)}+\delta^{ac}\delta^{bd}I_{\Delta,J}^{(0)}+\delta^{ab}\delta^{cd}I_{\Delta,J}^{(+)}
        \right)\,.
\end{equation}

To recover the conformal block decomposition is easy; for instance, take $z\in(0,1)$ and use the form of $\Psi_{\Delta,J}^{(0)}(z)$ given in Equation (\ref{eq:partialWaves0z1}). Then
\begin{equation}
    \Im(f^{abcd})(z) 
        = \sum_{J=0,1}\int_{\frac{1}{2}-i\infty}^{\frac{1}{2}+i\infty} \frac{d\Delta}{2\pi i} \frac{(-1)^J I^{abcd}_{\Delta,J}}{2K_{\Delta,J}} k_{\Delta}(z) 
        + \sum_{m=0}^{\infty} \frac{(-1)^m\Gamma(m)^2}{2\pi^2\Gamma(2m-1)} \tilde{I}^{abcd}_m k_m(z)\,, \label{eq:fourPointParialWave}
\end{equation}
and we can close the contour of the integral to the right of $\frac{1}{2}$ in the $\Delta$ plane to produce a sum over conformal blocks if we know the locations of poles of $\sum_{J} (-1)^J I_{\Delta,J}^{abcd}/2K_{\Delta,J}$. The location of a pole at $\Delta = \Delta_{\calO}$ denotes the dimension of an exchange operator $\calO$ in the OPE, and the residue is minus the coefficient in front of the conformal block (and, in principle, these are also three-point function coefficients squared, normalized by the two-point function coefficient, i.e. $-c_{\phi\phi\calO}^2/d_{\calO\calO}$).

We proceed with the analysis in pieces. $I^{(+)}_{\Delta,J}/2K_{\Delta,J}$ has no poles to the right of $\Delta=\frac{1}{2}$ so contributes nothing, hence we turn to $I^{(0)}_{\Delta,J}/2K_{\Delta,J}$. When $\Re(\alpha) \geq -\frac{1}{2}$ there are two infinite towers of poles which contribute, at $\Delta_{\calO} = \alpha+n+1$ and $\Delta_{\calO} = 2n+J$; the former come from poles of $I_{\Delta,J}^{(0)}$ and the latter from zeroes of $K_{\Delta,J}$. The residues are\footnote{To get from Equation (\ref{eq:residueEquation}) to Equation (\ref{eq:residueMassaged}) we need to use another identity from Sections 3.5 and 3.9 of \cite{bailey1935generalized}. In particular,
\begin{equation}
    \pFqreg{3}{2}{a,b,-n}{e,f}{1} 
        = (-1)^n \frac{\Gamma(1-f+b)\Gamma(1-f+a)}{\Gamma(e-c)\Gamma(f-c)} \pFqreg{3}{2}{1-s,1-f+c,-n}{1+b+c-f,1+a+c-f}{1}\,
\end{equation}
where $s=e+f-a-b-c$. In passing from Equation (\ref{eq:residueMassaged}) to Equation (\ref{eq:residueFinal}) we use our old identity from Equation (\ref{eq:coefficientIdentity}).}
\begin{align}
    \underset{\Delta = \alpha+n+1}{\mathbf{Res}} \frac{I^{(0)}_{\Delta,J}}{2K_{\Delta,J}} 
        &= (-1)^{n+J+1}\frac{\sqrt{\pi}\Gamma(-2\alpha-2n)}{2\Gamma(n+1)}\pFqreg{3}{2}{-\alpha-n,-\alpha-n,-n}{\frac{1}{2}-n,-2\alpha-2n}{1}\,,\label{eq:residueEquation}\\
        &= (-1)^{J+1}\frac{\Gamma(-2\alpha-2n)\Gamma(\alpha+n+1)^2}{2\Gamma(-2\alpha-n)\Gamma(n+1)}\pFqreg{3}{2}{\frac{1}{2},2\alpha+n+1,-n}{\alpha+1,\alpha+1}{1}\,,\label{eq:residueMassaged}\\
        &= (-1)^{J+1}\frac{1}{2}c_{\alpha+1,\frac{1}{2}}(n)\label{eq:residueFinal}\,;\\
    \underset{\Delta = 2n+J}{\mathbf{Res}} \frac{I^{(0)}_{\Delta,J}}{2K_{\Delta,J}} 
        &= (-1)^J\frac{4^{1-2n-J}\Gamma(2n+J)^3\Gamma(\alpha+2n+J)}{\pi \Gamma(2n+J-\frac{1}{2})}\label{eq:residueDiscreteCombined}\\
        &\qquad\times\,\pFqreg{3}{2}{2n+J,2n+J,\alpha+2n+J}{4n+2J,\alpha+2n+J+\frac{1}{2}}{1}\,.\nonumber
\end{align}

We can combine all of these results to obtain the residues of $\sum_J (-1)^{J} I_{\Delta,J}^{abcd}/2K_{\Delta,J}$ at $\Delta_{\calO} = \alpha + n + 1$. We have
\begin{align}
    \frac{\csc({\pi \alpha})}{\calN \calN_{\alpha}}\underset{\Delta = \alpha+n+1}{\mathbf{Res}} \sum_{J=0,1}\frac{(-1)^JI^{abcd}_{\Delta,J}}{2K_{\Delta,J}} 
        &= -\sum_{J=0,1} \frac{1}{2}c_{\alpha+1,\frac{1}{2}}(n) \left((-1)^{J}\delta^{ad}\delta^{bc}+\delta^{ac}\delta^{bd}\right)\,, \label{eq:4ptFromResidueUnsimplified}\\
        &= -\delta^{ac}\delta^{bd} c_{\alpha+1,\frac{1}{2}}(n)\,,\label{eq:4ptFromResidue}
\end{align}
which successfully reproduces the conformal block expansion. If instead we'd considered $z\in(-\infty,0)$ and the $\Psi^{(-)}(z)$ conformal partial waves, the calculations would proceed identically except Equation \eqref{eq:4ptFromResidueUnsimplified} would collect an extra factor of $(-1)^J$, projecting us onto the $\delta^{ad}\delta^{bc}$ sector.

Of course, we expect no physical operators at exchange dimensions $\Delta_{\calO}=2n$ and $\Delta_{\calO}=2n+1$ (atleast for $\Im(f^{abcd})$). As explained in \cite{Maldacena_2016, Mazac:2018qmi}, if we evaluate the contributions to the discrete series $\tilde{I}^{abcd}_m$ in Equation (\ref{eq:fourPointParialWave}), we get a cancellation with the spurrious residues at $\Delta_{\calO} = 2n$ and $\Delta_{\calO}=2n+1$. To illustrate the point, consider
\begin{align}
    \tilde{I}^{(0)}_{m} 
        &\coloneqq \int_{0}^1 \frac{dz}{z^2} \frac{z^{\alpha+1}}{\sqrt{1-z}}\Psi^{(0)}_{m, J}(z)\\
        &= 2\sqrt\pi (-1)^m \Gamma(m)^2 \Gamma(\alpha+m)\, \pFqreg{3}{2}{m,m,\alpha+m}{2m,\alpha+m+\frac{1}{2}}{1}\,,
\end{align}
then quantities such as
\begin{align}
    \frac{(-1)^m\Gamma(m)^2}{2\pi^2\Gamma(2m-1)} \tilde{I}^{(0)}_{m} 
        &= \frac{\Gamma(m)^4\Gamma(\alpha+m)}{\pi^{3/2}\Gamma(2m-1)}\pFqreg{3}{2}{m,m,\alpha+m}{2m,\alpha+m+\frac{1}{2}}{1}\,.
\end{align}
appear in the decomposition in Equation \eqref{eq:fourPointParialWave}. For $m = 2n$, this is Equation (\ref{eq:residueDiscreteCombined}) with $J=0$; for $m=2n+1$, this is negative Equation (\ref{eq:residueDiscreteCombined}) with $J=1$. By following the colour-index bookkeeping, we see that the discrete series contributions and the spurious poles cancel, as we anticipated!

Note the important role played by the $J=1$ eigenfunctions $\Psi_{\Delta,1}(z)$: if we had not included them, we would not have obtained the right result in Equation \eqref{eq:4ptFromResidue}. Instead, we would only have the colour-symmetric contribution, which is $\frac{1}{2}$ as large. This is familiar from the GFB case when we recall that $J=1$ is our $1d$ analogue of spin: the four-point function of four identical scalars involves the exchange of double trace operators with dimension $\Delta_{\calO} = 2\Delta_{\phi} + 2n$ (or $\Delta_{\calO} = 2\Delta_{\phi} + \ell+2m$ with spin $\ell\in 2\bbN^0$ in $d>1$). However, when colour is added, the colour-antisymmetric channel instead involves the exchange of operators with dimension $\Delta_{\calO} = 2\Delta_{\phi} + 2n + 1$.

Although there are a lot of similarities with the GFB and more conventional CFT, there are some differences. Consider the GFB, whose stripped four-point function is
\begin{equation}
    f_{\mathrm{GFB}}(z) = 1 + \abs{z}^{2\Delta_\phi} + \abs{\frac{z}{z-1}}^{2\Delta_\phi}\,.
\end{equation}
Before studying conformal partial waves, one must first remove the identity piece, $1$, as it's not normalizable with respect to the inner-product. However, because of the overall kinematic pre-factor $z/{\sqrt{1-z}}$, we seem to avoid this altogether, as all of the three pieces then have sensible conformal partial wave decompositions. 

The best analogue to a Generalized Free Theory, without the kinematic prefactor, actually comes from doing a celestial transform of the identity piece of a massive scattering amplitude \cite{ShuHengTalk}, i.e. the $\mathds{1}$ in $S = \mathds{1}+iT$. However, for massless scalars, the celestial transform of the identity piece only leads to contact terms unless some of the operators are at shadow weights of one another.

\subsection{Factorization}\label{sec:factorization}
As explained in Section \ref{sec:effDescON}, in the symmetry breaking phase of the $O(N)$ model, we are effectively scattering massless pions mediated by exchange of a massive $\sigma$-particle with some mass $M$ (we expand on this below). The CCFT three-point function coefficient for the scattering of two massless scalars with one massive scalar was computed in \cite{lamShao}, it is
\begin{equation}
    C_g(\Delta_1,\Delta_2;\Delta_{M}) = g \frac{M^{\Delta_1+\Delta_2-3}}{2^{\Delta_1+\Delta_2}} \frac{\Gamma\left(\frac{\Delta_1+\Delta_{M}-\Delta_2}{2}\right)\Gamma\left(\frac{\Delta_2+\Delta_{M}-\Delta_1}{2}\right)}{\Gamma(\Delta_{M})}\,,
\end{equation}
where $g$ is the coupling for the three-point interaction. Note: we have suppressed the colour indices in the following calculations, but it is straightforward to add them back in to the three-point functions by appending them with the appropriate $\delta^{ab}$ factors.

Meanwhile, the two point-function normalization is not ``1,'' because (in our conventions) the two point function is defined by the Mellin transform of a 2 particle amplitude in the original theory. We obtain the two-point function normalization for two massive particles $\sigma$ with mass $M$ by the same method as Appendix B of \cite{Atanasov:2021cje}. For a $d$-dimensional CCFT amplitude, the result is
\begin{equation}
    \expval{\sigma_{\Delta_1}(\vec{x}_1)\,\sigma_{\Delta_2}(\vec{x}_2)} 
        = \frac{2\pi^{d/2}}{M^{d}}\frac{\Gamma(\Delta_1-\frac{d}{2})}{\Gamma(\Delta_1)}\delta(\Delta_1-\Delta_2)\frac{1}{\abs{\vec{x}_1-\vec{x}_2}^{2\Delta_1}} \,.
\end{equation}
We will call the overall normalization $D_{\Delta_1\Delta_2}^M$.


Now, when the external particles are the same, the three-point function should square to the four point function (taking the two-point normalization into account). Recall that there were actually two poles on the second sheet of the $p^2$-plane, at $p_{\pm}$, associated with $\sigma$. We thus have two masses $M_{\pm}^2 := p_{\pm}^2$ in the $D_{\sigma\sigma}$ propagator and two couplings $g_{\pm}$ from Equation \eqref{eq:3point}. Using that
\begin{equation}
    p_{\pm} = -\sqrt{-2\mu^2} \, e^{\mp i\theta}\,,
\end{equation}
we have
\begin{align}
    M_{\pm}^2 
        &= (-2\mu^2) \, e^{\mp 2 i \theta}\,,\\
    g_{\pm}^2
        &= M_{\pm}^2\left(\frac{\lambda}{N} \csc\theta \right)(\mp i e^{\mp i \theta})\,.
\end{align}
Note: in Lorentzian signature these poles become $\omega_{\pm} = \sqrt{-2\mu^{2}} e^{-i \pi/2} e^{\mp i \theta}$.

And we can verify, at least for the ($\alpha$-independent) single trace tower in Equation \eqref{eq:conformalBlockTowers} with $\Delta_{M}=2n+1$, that
\begin{align}
    \sum_{s=\pm }\frac{C_{g_s}(\Delta_{\phi},\Delta_{\phi}; \Delta_{M_{s}})^2}{D_{\Delta_M\Delta_M}^{M_{s}}}
        &= \sum_{s=\pm} \pi g_s^2 \frac{M_{s}^{4\Delta_\phi-5}}{2^{4\Delta_\phi+1}} c_{1,\frac{1}{2}}(2n)\\
        &= 2\pi \left(\frac{\lambda}{N}\csc\theta\right) (-2\mu^2)^{2\Delta_{\phi}-\frac{3}{2}}\sin\Big(\left(2-4\Delta_{\phi}\right)\theta\Big) c_{1,\frac{1}{2}}(2n)\\
        &= \frac{\sin(4\pi \Delta_{\phi})}{2} \calN \calN_{\alpha} c_{1,\frac{1}{2}}(2n)\,. \label{eq:3PointFunctionSquares}
\end{align}
where $c_{p,q}(n)$ is defined in \eqref{eq:defcpq}. Two comments are in order. First, we see that the three-point functions squared, normalized, and summed reproduces the coefficients in the conformal block decomposition up to an overall $\Delta_{\phi}$ dependent factor. This is similar to the findings of \cite{Atanasov:2021cje} on integer mode exchanges in the conformal block decomposition of 4d massless scalars mediated by a massive exchange, and still requires explanation.

We also note the importance of summing over both resonances. If we had not summed over them we would not have been able to reproduce terms like $\sin((2-4\Delta_\phi)\theta)$ living inside the prefactor $\calN_{\alpha}$.

The double trace operators at $\Delta_{\calO} = \alpha+n+1$ aren't as straightforward. If we focus on just the $n$-dependent pieces of the four-point function coefficients, i.e. terms like 
\begin{align}
    c_{\alpha+1,\alpha+\frac{1}{2}}(n) 
        &\coloneqq \frac{(\alpha+1)^2_n}{n! (2\alpha+n+1)_n} \pFq{3}{2}{\frac{1}{2},2\alpha+n+1,-n}{\alpha+1,\alpha+1}{1}\,,\\
        &=2^{-2(\alpha+n)}\frac{\Gamma(\alpha+n+1)}{\Gamma(\alpha+n+\frac{1}{2})} \sum_{k=0}^n  \frac{(-1)^k\Gamma(k+\frac{1}{2})}{\Gamma(k+1)\Gamma(\alpha+k+1)^2}\frac{\Gamma(2\alpha+n+k+1)}{\Gamma(n-k+1)}
\end{align}
then it's easy to see, by comparing to the $n$-dependence of our previous guess, that something of the form
\begin{equation}
    \frac{C_{g}(\Delta_{\phi},\Delta_{\phi}; \Delta_{\calO})^2}{D_{\Delta_{\calO}\Delta_{\calO}}^{M}}
\end{equation}
will not match, although the expansion of $c_{\alpha+1,\alpha+\frac{1}{2}}(n)$ has a very suggestive form.

There are a couple possible explanations for this non-matching of the three-point functions to the four-point function coefficients. Some of them include:
\begin{enumerate}
    \item There are multiple operators with the same scaling dimension $\Delta_{\calO} = \alpha+n+1$, so one cannot just naively identify the coefficients of a four-point function with one ``three-point function squared.''
    \item There are other three-point functions in the CCFT that do not come from the celestial transform of a three-particle amplitude. A straightforward conjecture is that the ``double trace operators'' come from scattering two-particle states.
    \item CCFTs are different from standard CFTs.
\end{enumerate}

Finally, in \cite{lamShao}, the authors show that the optical theorem for the scattering amplitude translates into a decomposition of $\Im(f(z))$ along the principal continuous series partial waves. In the case of the $O(N)$ amplitude (or more broadly, any amplitude with resonances), this is modified in the same way as our previous discussions, in that one must sum over resonances to obtain the analogous formula, i.e.
\begin{equation}
    \Im f(z) = \pi \sum_{s=\pm} M_{\pm} \int_{-\infty}^{\infty} d\nu\,\mu(\nu) C_{g_s}\left(\Delta_1,\Delta_2; \frac{1}{2}+i\nu\right)C_{g_s}\left(\Delta_3,\Delta_4; \frac{1}{2}-i\nu\right) \Psi_{\frac{1}{2}+i\nu}(z)\,.
\end{equation}
This follows directly from applying the same manipulations in \cite{lamShao} to Equation \eqref{eq:AmplitudeFactorization}, since Equation \eqref{eq:AmplitudeFactorization} shows exactly how the four-point celestial amplitude is writable as a (sum of) three-point amplitude(s) squared, and (aside from the fact that there are two terms) takes the exact same form as the amplitude in the reference.\footnote{Another more physical way to see this follows from the results in Section 6 of \cite{Melton:2021kkz}. In particular, the author shows how an amplitude which can be factorized into two pieces, connected by the exchange of an off-shell massive particle, can be written as an integral over conformal partial waves with an additional integral over internal momentum of the exchange particle weighted by a Feynman propagator. Using the usual Cutkosy rules on this line (replacing the Feynman propagator with a delta function) so that the exchange particle is on-shell, one obtains the celestial optical theorem from \cite{lamShao}. This will be explained in more detail in forthcoming work.}

        



		

\section{Conclusion}\label{sec:conclusion}
In this paper we studied the effects of unstable particles on unitarity of scattering amplitudes (via the optical theorem) when phrased in variables on the celestial circle (or the celestial sphere more generally), and the effects on factorization of celestial amplitudes and the celestial optical theorem. We also discussed the 3d $O(N)$ model as an interesting example for celestial holography in it's own right.

In particular, we studied how a $2\to 2$ scattering amplitude written in terms of Mandelstam variables could be re-written in terms of center of mass energy $\omega$ and a coordinate $z$ on the celestial circle. In this basis, the poles of the S-matrix ``unfold'' into poles in the complex $\omega$-plane, where their effects on the optical theorem manifest geometrically as whether or not they are caught in simple contour integrals. For example, an unstable particle (a pole on the second sheet in Mandelstam variables) becomes a pole in the lower-half $\omega$ plane, which is then turned into a pole in the upper-half plane in the imaginary amplitude and caught by a contour integral when doing a Mellin transform. Our discussion also points at inherently celestial CFT ways to identify unstable particles in scattering amplitudes. We leave development of a complex-analytic description of celestial amplitudes (in the spirit of the S-matrix bootstrap program) to future work.

As a motivating example, we applied our results to the 3d $O(N)$ model and worked with the fully resummed $\pi\pi\to \pi\pi$ scattering amplitude in the large $N$ limit. The 3d $O(N)$ model is interesting as it is both UV and IR finite, satisfies positivity constraints in the EFT expansion, and contains unstable states in the spectrum. Moreover, taking the large $N$ limit allowed us to study the theory at all loop orders and finite coupling. From the CCFT point-of-view, this gives an interesting example of an interacting non-supersymmetric celestial amplitude beyond tree-level, which could be a useful example for understanding both renormalization and spontaneous symmetry breaking in CCFT; we also leave these studies to future investigations. When converted to a celestial amplitude, this model showcased a number of interesting features:
\begin{enumerate}
    \item The resummed amplitude shifted the convergence strip for the celestial scaling dimension $\Delta_T$ (also called $\alpha$ or $\beta$ in the literature) in the Mellin transform to include the principal continuous series.
    \item A CCFT four-point function analogous to a Generalized Free Field Theory.
    \item Confirmed that scattering of bulk pions with ``Bose statistics'' behave like a correlation function of bosons in the CCFT.
    \item The effects of unstable particles and/or resonances on CCFT amplitudes.
\end{enumerate}

To elaborate on the last point. In the final section of the paper we provided independent conformal block and conformal partial wave decompositions of the $\pi\pi\to\pi\pi$ amplitude; and verified they were equivalent after a standard cancellation of spurious poles along the principal discrete series. In particular, we found that there were two towers of exchanged particles in these conformal decompositions, one exchanging ``single trace'' states, and one exchanging ``double trace'' states. On one hand, we found that the single trace tower was able to be given an interpretation as the exchange of massive particles with increasing conformal dimension. However, we also found that it was essential to interpret the mass of the particles to be coming from both of the massive unstable resonances when doing so, or else one would not be able to produce the correct prefactors (with the correct dependence on external scaling dimensions and couplings). On the other hand, the double trace towers did not factorize, and we conjectured some reasons for this; we leave the resolution of this to future work. Finally, we comment on the CCFT realization of the optical theorem, and explain how it must also be modified to include a sum over resonances in the case of the $O(N)$ model, for the exact same reasoning as in the conformal block decomposition; we also leave a more detailed description of cutting rules in CCFT to future work.

\subsection{Open Problems}\label{sec:openProblems}
In this section, we enumerate a few more speculative and conjectural directions spurred on by our investigations, but not necessarily with an immediate connection to the main text.
\begin{enumerate}
    \item \textbf{Other large $N$ theories}. In general, the large $N$ limit allows one to go beyond tree level and compute observables exactly as functions of the coupling constants, providing a standard approach to non-perturbative physics. In our case, this led to an interesting shift of the convergence region for the Mellin integrals, amongst other properties. There are many other interesting theories which might be elucidated by a celestial analysis, or conversely, provide interesting case studies to probe the CCFT formalism. One example might be the Gross-Neveu model \cite{Gross:1974jv}; for (2+1)d Gross-Neveu it would be satisfying to see relationships to Generalized Free Fermions or SYK emerge (see also \cite{Pasterski:2022joy,Kar:2022vqy}).
    
    \item \textbf{Reversing the celestial transform}. In practice, we typically start with a scattering amplitude in $(d+2)$-dimensions then turn it into the correlator of a $d$-dimensional CCFT. It could be interesting to see if there are any structures about $(d+2)$-dimensional amplitudes that are revealed by reversing this procedure. For example, the CCFT correlation function obtained from the $O(N)$ model roughly looks like a modification of a generalized free boson by a ``kinematic pre-factor'' and an exponential related to the scattering channel. Can one start with a (possibly modified) CFT correlation function, e.g. for a 1d generalized free fermion or the SYK model, and see what (2+1)d amplitudes it produces? Since we are currently missing an intrinsic definition of a CCFT, it may be hard to justify any starting correlation function a priori.
    
    \hspace{0.5cm} In a similar vein, are there any ``manipulations'' that can be performed on a CCFT that become new (or recover known) dualities of scattering amplitudes when both the starting and ending CCFT are reverse-transformed? If a duality is apparent at the level of S-matrix elements (e.g. as in Chern-Simons Matter theories \cite{Jain:2014nza}), then this should have some manifestation in CCFT correlation functions. It would be curious to see if whatever relationship appears could be abstracted though to find new dualities.
    
    \hspace{0.5cm} More generally we are asking to what extent we can commute around diagrams like
    \begin{equation}
    \begin{tikzcd}
    	{\mathcal{A}_1} & {\mathcal{A}_2} \\
    	{\mathrm{CCFT}_1} & {\mathrm{CCFT}_2}
    	\arrow[tail reversed, from=1-2, to=2-2]
    	\arrow[from=2-1, to=2-2]
    	\arrow[from=1-1, to=1-2]
    	\arrow[tail reversed, from=1-1, to=2-1]
    \end{tikzcd}
    \end{equation}
    and track interesting properties of QFTs like dualities, RG flows, etc., with the goal of applying CCFT formalism to understanding QFT more broadly.
    
    \item \textbf{Theories with Anyons}. In Section \ref{sec:SMatCrossing} we pointed out that a naive permutation symmetry of a (2+1)d scattering amplitude led to the condition that correlation functions of the CCFT behaved ``like bosons.'' One might conjecture that given an S-matrix in a theory such as those in \cite{Jain:2014nza}, where particles are dressed with Wilson lines with a Chern-Simons part of the action, that some non-trivial features of the S-matrix could be revealed.
    
    \item \textbf{CCFTs and Discrete Symmetries}. Discrete actions on $(d+2)$-dimensional Minkowski spacetime induce discrete actions on the $d$-dimensional celestial circle. For example, in (2+1)d a generic scattering amplitude can depend on the orientation of the scattering momenta (see \cite{Jain:2014nza} for example). The non-invariance of such a scattering amplitude under reflections should imply non-invariance of the 1d CCFT under orientation-reversal.
    
    \item \textbf{Lorentzian inversion formula and dispersion relations}. Another way to obtain the coefficients $I_{\Delta,J}$ is given by the Lorentzian inversion formula \cite{Caron_Huot_2017,Simmons_Duffin_2018} (see also \cite{Maldacena_2016}). It is easier in that one only needs to integrate against the $\dDisc$, defined by
    \begin{equation}
        \dDisc[f(z)] \coloneqq f_t(z) - \frac{1}{2} \left( f_s^{\aboveContour}(z) +f_s^{\belowContour}(z) \right)\,, \qquad z \in (0,1)\,,
    \end{equation}
    where $f_s^{\aboveContour}(z)$ and $f_s^{\belowContour}(z)$ are the continuations of $f(z)$ from the $z\in(1,\infty)$ region to the $z\in(0,1)$ region above or below the real axis. 
    
    As an example, for our $O(N)$ model the negative average is
    \begin{equation}
        - \frac{1}{2} \left( f_s^{\aboveContour}(z) +f_s^{\belowContour}(z) \right)     = \delta^{ad}\delta^{bc} \sin(\pi\alpha) \frac{z}{\sqrt{1-z}}\left(\frac{z}{1-z}\right)^{\alpha}\,,
    \end{equation}
    so that $\dDisc[f(z)]$ in $z\in(0,1)$ is just
    \begin{equation}
        \dDisc[f(z)] = 
            \calN \calN_{\alpha} \frac{z}{\sqrt{1-z}} \left(\delta^{ab}\delta^{cd}+\delta^{ac}\delta^{bd}e^{i\pi\alpha}z^{\alpha}+\delta^{ad}\delta^{bc}(1+\sin(\pi\alpha))\left(\frac{z}{1-z}\right)^\alpha\right)\,.
    \end{equation}
    
    \hspace{0.5cm} Part of the elegance of the Lorentzian inversion formula is that it gives $I_{\Delta,J}$ in a way analytic in spin, in addition to various analytic bounds on OPE data and diagnostics for chaos (see \cite{Maldacena_2016, Mazac:2018qmi} and references within). In $d=1$ there is no analyticity in spin, but analyticity in the discrete series $I_{m,J}$ \cite{Simmons_Duffin_2018}.
    
    \hspace{0.5cm} Since analytic continuation in $z$ relates one scattering channel to another \cite{lamShao, Chang:2021wvv}, it could be interesting to understand which data of scattering amplitudes can be extracted from understanding $\dDisc$'s in the CCFT.
    
    
    \item \textbf{2d Bulks}. In this paper we studied (2+1)d amplitudes which became correlation functions on the celestial circle. We might try to study (1+1)d bulks so that the boundary theory is supported entirely on two points, leaving almost all physics to the $\Delta_T$-plane (or $\alpha$ or $\beta$ plane in other conventions). It could be interesting to transform the Sine-Gordon theory to these ``celestial points,'' or other integrable theories, to see how interesting non-perturbative phenomena manifest in the celestial presentation.
\end{enumerate}

\acknowledgments
We would like to thank Davide Gaiotto for his comments throughout the research process. We would like to thank Davide Gaiotto and Sabrina Pasterski for feedback on the draft. AG thanks Shu-Heng Shao and Andy Strominger for discussions, as well as the organizers and participants of the "S-Matrix Bootstrap 2022" conference for hospitality and valuable exchange of ideas. 

The work of DGS is supported by a Plotnick Fellowship in the Enrico Fermi Institute at the University of Chicago and in part by the US Department of Energy DE-SC0021432. JK is funded through the NSERC CGS-D program. JW is supported in part by a grant from the Krembil foundation by the Perimeter Institute for Theoretical Physics. AG was partially supported by DOE
grant de-sc/000787, by the Society of Fellows and by the Black Hole Initiative at Harvard University, which is funded by grants from the John
Templeton Foundation and the Gordon and Betty Moore
Foundation. Research at Perimeter Institute is supported in part by the Government of Canada through the Department of Innovation, Science and Economic Development Canada and by the Province of Ontario through the Ministry of Colleges and Universities. JW is also supported by the European Union's Horizon 2020 Framework: ERC grant 682608 and the "Simons Collaboration on Special Holonomy in Geometry, Analysis and Physics'

\appendix
\section{CFT Background and Conventions}\label{app:CFTBackgroundConventions}
Here we record background and references on CFT, general reviews of CFT are \cite{Rychkov:2016iqz, simmonsduffin2016tasi}. We often specialize to 1d where things are slightly different from other dimensions. An axiomatic approach to \textit{unitary} 1d CFT appears in \cite{Qiao:2017xif}, but some lessons are transmutable; helpful reviews of 1d conformal partial waves appear in \cite{gadde2017search, Mazac:2018qmi, Hogervorst:2021uvp}.

\subsection{Four-Point Functions in 1d CFT}\label{sec:1dCFTGeneral}
In a 1d CFT, as in higher dimensions, a four-point correlation function is fixed by conformal invariance to take the form
\begin{equation}
    \expval{\phi_1^a(x_1)\phi_2^b(x_2)\phi_3^c(x_3)\phi_4^d(x_4)} = \frac{\left(\frac{x_{14}^2}{x_{24}^2}\right)^{-\frac{1}{2}\Delta_{12}}\left(\frac{x_{14}^2}{x_{13}^2}\right)^{\frac{1}{2}\Delta_{34}}}{\left(x_{12}^2\right)^{\frac{1}{2}(\Delta_1+\Delta_2)}\left(x_{34}^2\right)^{\frac{1}{2}(\Delta_3+\Delta_4)}} f^{abcd}(z) \label{eq:4ptfunc}\,,
\end{equation}
where $\Delta_{ij} \coloneqq \Delta_i - \Delta_j$ and the conformal cross-ratio is
\begin{equation}
    z \coloneqq \frac{x_{12} x_{34}}{x_{13} x_{24}}\,.
\end{equation}
If we take all $\phi_i$ to have the same conformal weight $\Delta_\phi$, then this becomes
\begin{equation}
    \expval{\phi^a(x_1)\phi^b(x_2)\phi^c(x_3)\phi^d(x_4)} 
        = \frac{1}{x_{12}^{2\Delta_\phi}x_{34}^{2\Delta_\phi}} f^{abcd}(z)\,.\label{eq:4ptfuncSameWeight}
\end{equation}
We will also call (stripped) four-point functions, like $f^{abcd}(z)$, ``four-point functions.''


It is convenient for intuition to spend the $SL(2,\bbR)$ symmetry and set $(x_1,x_2,x_3,x_4)=(0,z,1,\infty)$. Since the four point function is singular at $z=0$, $1$, $\infty$, the 1d CFT four-point function is really only defined piecewise. We suggestively label the three disconnected regions with $u$, $t$, and $s$, anticipating applications to CCFT, so that
\begin{equation}
    f^{abcd}(z) \coloneqq
        \begin{cases}
            f_u^{abcd}(z) & \text{if $z\in(-\infty,0)$}\\
            f_t^{abcd}(z) & \text{if $z\in(0,1)$}\\
            f_s^{abcd}(z) & \text{if $z\in(1,\infty)$}
        \end{cases}\,.
\end{equation}
As explained in Section 2 of \cite{Mazac:2018qmi}, these $f^{abcd}_{u,t,s}(z)$ can be analytically continued to other regions, but they are not necessarily the analytic continuations of one another.\footnote{Although in the specific case of 1d CCFT correlation functions, they are actually related in a slightly non-trivial way, see e.g. \cite{lamShao, Chang:2021wvv}. \label{footnote:ChangPaper}} 

However, if we started with a correlation function of four bosons, then various $f^{abcd}(z)$ are related by relabelling. For instance, invariance under relabelling $1 \leftrightarrow 2$ (equivalently $3 \leftrightarrow 4$) and $2 \leftrightarrow 3$ (or $1 \leftrightarrow 4$) imply
\begin{alignat}{3}
    f_t^{abcd}\left(\frac{z}{z-1}\right) 
        &= f_u^{bacd}(z) = f_u^{abdc}(z)\,, 
        && \quad z\in(-\infty,0)\,,\label{eq:commRln1}\\
    z^{2\Delta_{\phi}} f_t^{abcd}\left(\frac{1}{z}\right) 
        &= f_s^{acbd}(z) = f_s^{dbca}(z)\,, 
        && \quad z\in(1,\infty)\,,\label{eq:commRln2}
\end{alignat}
respectively. Moreover, swapping $1\leftrightarrow 3$ (or $2\leftrightarrow 4$) gives CFT crossing symmetry
\begin{equation}
    z^{-2\Delta_{\phi}}f_t^{abcd}(z)
        = (1-z)^{-2\Delta_{\phi}} f_t^{adcb}(1-z) 
        = (1-z)^{-2\Delta_{\phi}} f_t^{cbad}(1-z)\,, 
         \quad z\in(0,1)\,.\label{eq:commRln3}\\
\end{equation}
Similar relations exist for fermionic statistics \cite{Mazac:2018qmi}. Moreover, care should be taken about some transformations inducing time-reversal when it is a concern \cite{Bulycheva_2017}. These CFT crossing symmetry constraints (including the additional 1d relations) are familiar and heavily used in, for example, the conformal bootstrap program (see e.g. \cite{Rattazzi:2008pe, Mazac:2018qmi, El-Showk:2012cjh, Kos:2013tga, Poland:2018epd, Ferrero:2019luz, He:2020azu} for various references including recent results in 1d). We expand a bit more on this in the next section.

\subsection{Decomposing CFT Four-Point Functions}\label{app:blocksAndWaves}
\subsubsection{Conformal Blocks}\label{app:confBlock}
There are 3 different channel decompositions for $f(z)$ that can be performed using the OPE, but the convergence of the various OPEs depends on the value of $z$ (which translates to physical region for the CCFT scattering amplitude). In particular, when $z\in(0,1)$ the expansion where $z\to0$ (contracting $\phi_1\phi_2$ and $\phi_3\phi_4$) converges, telling us\footnote{This $z\to0$ contraction is sometimes called the $s$-channel OPE, while the $z\to1$ contraction is called $t$-channel OPE, and the $z\to\infty$ contraction is called $u$-channel OPE. We will avoid overloading the terms $s$, $t$, and $u$ any further.}
\begin{equation}
    f(z) = \sum_{\calO} \frac{c_{\phi\phi\calO}^2}{d_{\calO\calO}} k^{}_{\Delta_{\calO}}(z)\,.
\end{equation}
Here $\calO$ denotes an exchanged conformal primary in the $\phi\times\phi$ OPE with scaling dimension $\Delta_{\calO}$, and $k^{}_{\Delta_{\calO}}(z)$ are the $SL(2,\bbR)$ conformal blocks, proto-correlators that span the vector space of solutions to the Ward identities. The coefficients $c_{\phi\phi\calO}$ are the usual three point function coefficients, and $d_{\calO\calO}$ is the normalization of the two-point function. In a unitary theory, the $d_{\calO\calO}$ are typically chosen to be $1$ (see \cite{Rychkov:2016iqz}), but in CCFT we do not use this freedom, opting to define the two-point function from the celestial transform.

In less obscure terms, the conformal blocks are functions which re-sum over all descendants of the primary $\calO$, and hence describe the contribution to $\expval{\phi_1\phi_2\phi_3\phi_4}$ by exchange of $\calO$ between $\phi_1\phi_2$ and $\phi_3\phi_4$ \cite{Dolan:2003hv,Dolan:2011dv}. They are known to have a nice closed form in terms of hypergeometric functions in $d=1$,$2$,$4$. Focusing on $d=1$, the $SL(2,\bbR)$ conformal blocks are
\begin{equation}
    k^{\Delta_i}_{\Delta}(z) = z^\Delta {}_2F_1(\Delta-\Delta_{12},\Delta+\Delta_{34};2\Delta;z)\,. \label{eq:SL2Rblock}
\end{equation}
Expanding in other channels, and enforcing associativity of OPE, is what gives crossing symmetry in the CFT.

\subsubsection{Conformal Partial Waves}\label{sec:CPWFormulas}
Another related expansion is the ``conformal partial wave expansion,'' which is an expansion in a complete basis of single-valued functions $\Psi_{\Delta,J}(z,\bar{z})$ called conformal partial waves, which are analogs of harmonic functions for the conformal group \cite{gadde2017search}. They are typically chosen to be the eigenfunctions of the $z\to 0$ channel Casimir.

For general $d>1$, the so-called ``principal series representations'' consisting of those eigenfunctions having $\Delta=\frac{d}{2}+ir$  with $r\in\bbR_{\geq0}$, and integral spin $J$, form a complete and orthogonal set of $\Psi$ with respect to an inner product $(\,\,\,,\,\,\,)$.\footnote{For brevity we suppress the details of the integral defining the inner product in arbitrary dimensions, see \cite{Simmons_Duffin_2018} for details.} The moral of the story is that this allows one to write
\begin{equation}
    f(z,\bar{z}) 
        = \sum_{J=0}^\infty \int_{\frac{d}{2}}^{\frac{d}{2}+i\infty} \frac{d\Delta}{2\pi i} \frac{I_{\Delta,J}}{n_{\Delta,J}} \Psi_{\Delta,J}(z,\bar{z}) + (\textrm{non-norm.}),
\end{equation}
where $I_{\Delta,J}$ contains the dynamical data of the CFT (i.e. the data in three-point functions) and can be naively extracted using the aforementioned inner product; this is called the Euclidean inversion formula. Or, famously, extracted in a way analytic in $J$ using the Lorentzian inversion formula \cite{Caron_Huot_2017, Simmons_Duffin_2018}.

Specializing to $d=1$, we also need to include the ``principal discrete series'' to form a complete set of partial waves. Altogether we have a basis consisting of:\footnote{We will also use the notation $\Psi_{m,J}$ for the discrete series wavefunctions, keeping in mind that $\Psi_{m,J}$ is only normalizable (and thus relevant for us) when $J \equiv m \Mod 2$.}
\begin{alignat}{3}
    &\Psi_{\Delta,J}\,,  &&\quad \Delta\in \frac{1}{2}+i\bbR_{\geq0}\,,\\
    &\Psi_{m,0}\,,       &&\quad m\in 2\bbN\,,\\
    &\Psi_{m,1}\,,       &&\quad m\in 2\bbN-1\,,
\end{alignat}
where $J=0,1$ is a 1d analog of spin, and denotes if $\Psi_{\Delta,J}$ is symmetric or anti-symmetric respectively under $z\mapsto \frac{z}{z-1}$. The inner product is given by
\begin{equation}
    (g,f) = \int_{-\infty}^{\infty} \frac{dz}{z^2} \bar{g}(z) f(z)\,,
\end{equation}
and the eigenfunctions satisfy
\begin{align}
    \left(\Psi_{\frac{1}{2}+ir,J},\Psi_{\frac{1}{2}+ir^{\prime},J^\prime}\right)
        &= 2\pi n_{\frac{1}{2}+ir} \delta(r-r^\prime) \delta_{J,J^\prime}\,,\\
    \left(\Psi_{m,J},\Psi_{n,J^\prime}\right)
        &= \frac{4\pi^2}{2m-1} \delta_{m,n} \delta_{J,J^\prime}\,,\\
    \left(\Psi_{\frac{1}{2}+ir,J},\Psi_{n,J^\prime}\right)
        &= 0\,.
\end{align}
Here we define
\begin{equation}
    K_{\Delta,J} \coloneqq \frac{\sqrt{\pi}\Gamma(\Delta-\frac{1}{2})\Gamma(\Delta-1+J)\Gamma\left(\frac{1}{2}(1-\Delta+J)\right)^2}{\Gamma(\Delta-1)\Gamma(1-\Delta+J)\Gamma\left(\frac{\Delta+J}{2}\right)^2}\,,
\end{equation}
\begin{equation}
    n_{\Delta} \coloneqq 2K_{\Delta,J}K_{1-\Delta,J} = \frac{4\pi \tan(\pi\Delta)}{2\Delta-1}\,.
\end{equation}
So that in total we can write
\begin{equation}
    f(z) 
        = \sum_{J} \int_{\frac{1}{2}}^{\frac{1}{2}+i\infty} \frac{d\Delta}{2\pi i} \frac{I_{\Delta,J}}{n_{\Delta}} \Psi_{\Delta,J}(z) 
        + \sum_{m=1}^{\infty} \frac{2m-1}{4\pi^2} \tilde{I}_{m} \Psi_{m,J}(z)\,,\label{eq:1dpartialWaveDecomposition}
\end{equation}
recalling again that for the discrete series $J \equiv m \Mod 2$. 

The explicit formulas for the 1d conformal partial waves are
\begin{align}
    \Psi_{\Delta,0}(z) 
        &= \int dx_5\, \left(\frac{\abs{x_{12}}}{\abs{x_{15}}\abs{x_{25}}}\right)^{\Delta} \left(\frac{\abs{x_{34}}}{\abs{x_{35}}\abs{x_{45}}}\right)^{1-\Delta}\,,\\
    \Psi_{\Delta,1}(z) 
        &= \int dx_5\, \left(\frac{\abs{x_{12}}}{\abs{x_{15}}\abs{x_{25}}}\right)^{\Delta} \left(\frac{\abs{x_{34}}}{\abs{x_{35}}\abs{x_{45}}}\right)^{1-\Delta}\mathrm{sgn}(x_{12}x_{15}x_{25}x_{34}x_{35}x_{45})\,.
\end{align}
The integrals converge when $0 < \Re(\Delta) < 1$, and other values can be obtained by analytic continuation from these answers. One can also evaluate these integrals explicitly in the three different regions for $z$. For the principal continuous series we have\footnote{We will use the labels $(-)$, $(0)$, and $(+)$ for the conformal partial waves in the regions $(-\infty,0)$, $(0,1)$, and $(1,\infty)$ respectively, to avoid deviations of notation from the conformal partial wave literature. For example, $f_t(z)$ is decomposed along the $\Psi^{(0)}$ conformal partial waves.}
\begin{equation}
    \Psi_{\Delta,J}^{(0)}(z) = (-1)^J K_{1-\Delta,J} \, k_{\Delta}(z)+ (-1)^J K_{\Delta,J} \, k_{1-\Delta}(z)\,,\qquad z\in (0,1)\,, \label{eq:partialWaves0z1}
\end{equation}
and for the principal discrete series
\begin{equation}
    \Psi_{\Delta,J}^{(0)}(z) = (-1)^J K_{1-\Delta,J} \, k_{\Delta}(z)\,,\qquad z\in (0,1)\,.
\end{equation}
We can write the conformal partial wave in the other two regions in terms of $\Psi_{\Delta,J}^{(0)}$ as
\begin{alignat}{3}
    \Psi_{\Delta,J}^{(-)}(z)
        &= (-1)^J \Psi_{\Delta,J}^{(0)}\left(\frac{z}{z-1}\right)\,,\qquad && z\in(-\infty,0)\,,\label{eq:PsiMinusFromZero}\\
    \Psi_{\Delta,J}^{(+)}(z)
        &= \frac{1}{2}\left[\Psi_{\Delta,J}^{(0)}(z+i\epsilon)+\Psi_{\Delta,J}^{(0)}(z-i\epsilon)\right]\,,\qquad && z\in(1,\infty)\,.
\end{alignat}
It can be helpful to know that this last piece evaluates to
\begin{equation}
    \Psi_{\Delta,J}^{(+)}(z)
        = \frac{2^{J+1}}{\sqrt{\pi}}\left(\frac{z-2}{z}\right)^{\delta_{1J}} \Gamma\left(\tfrac{1-\Delta+J}{2}\right)\Gamma\left(\tfrac{\Delta+J}{2}\right) \,{}_2F_{1}\left(\tfrac{1-\Delta+J}{2},\tfrac{\Delta+J}{2};\tfrac{1}{2}+J;\tfrac{(2-z)^2}{z^2}\right)\,.
\end{equation}
For further technical details see results in \cite{Maldacena_2016, Bulycheva_2017, Simmons_Duffin_2018, Mazac:2018qmi, Hogervorst:2021uvp}.

\bibliographystyle{JHEP}
\bibliography{CelestialON}

\end{document}